\documentclass[authoryear,preprint,review,12pt]{elsarticle}

\usepackage{graphics}
\usepackage{natbib}
\usepackage{setspace}
\usepackage{amssymb}
\usepackage{amstext}
\usepackage{textcmds}
\usepackage{color}
\usepackage{longtable}
\usepackage{tabularx}
\usepackage{array}
\usepackage[version=3]{mhchem}

\RequirePackage{hyperref}
\hypersetup{colorlinks=true, citecolor=black, breaklinks=true, filecolor=black, linkcolor=black, urlcolor=black, bookmarks=true}

\small\normalsize

\journal{Icarus}

\begin{document}

\begin{frontmatter}

\title{2D photochemical modeling of Saturn's stratosphere \\ Part I: Seasonal variation of atmospheric composition without meridional transport}

\author[LAB1,LAB2]{V. Hue\corref{cor1}} \ead{Vincent.Hue$@$obs.u-bordeaux1.fr} 
\author[MPS]{T. Cavali\'e}
\author[LAB1,LAB2]{M. Dobrijevic}
\author[LAB1,LAB2]{F. Hersant}
\author[SwRI]{T. K. Greathouse}

\address[LAB1]{Universit\'e de Bordeaux, Laboratoire d'Astrophysique de Bordeaux, UMR 5804, F-33270 Floirac, France}
\address[LAB2]{CNRS, Laboratoire d'Astrophysique de Bordeaux, UMR 5804, F-33270, Floirac, France}
\address[MPS]{Max-Planck-Institut f\"ur Sonnensystemforschung, 37077, G\"ottingen, Germany}
\address[SwRI]{Southwest Research Institute, San Antonio, TX 78228, United States}

\cortext[cor1]{Tel: +33-5-5777-\textbf{6164}}

\begin{abstract}

Saturn's axial tilt of 26.7$^{\circ}$ produces seasons in a similar way as on Earth. Both the stratospheric temperature and composition are affected by this latitudinally varying insolation along Saturn's orbital path. A new time-dependent 2D photochemical model is presented to study the seasonal evolution of Saturn's stratospheric composition. This study focuses on the impact of the seasonally variable thermal field on the main stratospheric C2-hydrocarbon chemistry (C$_2$H$_2$ and C$_2$H$_6$) using a realistic radiative climate model. Meridional mixing and advective processes are implemented in the model but turned off in the present study for the sake of simplicity. The results are compared to a simple study case where a latitudinally and temporally steady thermal field is assumed. Our simulations suggest that, when the seasonally variable thermal field is accounted for, the downward diffusion of the seasonally produced hydrocarbons is faster due to the seasonal compression of the atmospheric column during winter. This effect increases with increasing latitudes which experience the most important thermal changes in the course of the seasons. The seasonal variability of C$_2$H$_2$ and C$_2$H$_6$ therefore persists at higher-pressure levels with a seasonally-variable thermal field. Cassini limb-observations of C$_2$H$_2$ and C$_2$H$_6$ \citep{Guerlet2009} are reasonably well-reproduced from the equator to 40$^{\circ}$ in both hemispheres in the 0.1-1\,mbar pressure range. At lower pressure levels, the models only fit the Cassini observations in the northern hemisphere, from the equator to 40$^{\circ}$N. Beyond 40$^{\circ}$ in both hemispheres, deviations from the pure photochemical predictions, mostly in the southern hemisphere, suggest the presence of large-scale stratospheric dynamics.

\end{abstract}

\begin{keyword}
\textbf{Photochemistry \sep Saturn \sep Atmosphere, evolution}
\end{keyword}

\end{frontmatter}

\section{Introduction}

Observations of Saturn in the infrared and millimetric range, performed by ISO or ground-based facilities gave us access to its disk-averaged stratospheric composition (see the review of \citet{Fouchet2009} for a complete list of observations), for which 1D photochemical models have done a fairly good job reproducing it \citep{Moses2000a, Moses2000b}.
Close-up observations, performed by the Voyager missions as well as recent ground-based observations, have unveiled variations with latitude of the temperature and the stratospheric composition \citep{Ollivier2000a, Greathouse2005, Sinclair2014}.
The Cassini probe has now mapped (as a function of altitude and latitude) and monitored for almost 10 years, i.e. 1.5 Saturn season, the temperature and the main hydrocarbon emissions in Saturn's stratosphere \citep{Howett2007,Fouchet2008,Hesman2009,Guerlet2009,Guerlet2010,Li2010, Fletcher2010,Sinclair2013,Sinclair2014}.

We now have an impressive amount of data for which 1D photochemical models (e.g., \citealt{Moses2000a,Moses2000b,Moses2005a} and \citealt{Ollivier2000b}) have become insufficient in predicting the 3D properties of Saturn's stratosphere, especially in terms of dynamics (diffusion and advection). On the other hand, general circulation models (GCM) are being developed for Jupiter \citep{Medvedev2013} and Saturn \citep{Dowling2006, Dowling2010, Friedson2012, Guerlet2014}. Such models usually focus on dynamics and therefore are restricted in their description of the atmospheric chemistry as they are limited to only a few reactions, if any at all.

\citet{Liang2005} and \citet{Moses2005b} made the first attempts to construct latitude-altitude photochemical models for the giant planets, followed by \citet{Moses2007} who built a 2D-photochemical model for Saturn and who accounted for simple Hadley-type circulation cells as well as meridional diffusive transport. The quasi-two-dimensional model developed by \citet{Liang2005} does not fully account for the latitudinal transport as a diffusive correction is added at the end of the one-dimensional calculations. This model also does not account for evolution of the orbital parameters. Due to its very low obliquity, the seasonal effects on Jupiter should be mainly caused by its eccentricity and might be non negligible. On the other hand, the model developed by \citet{Moses2005b} accounts for the seasonal evolution of the orbital parameters as well as the variations in solar conditions. They have shown that, for Saturn, the seasonal effects on atmospheric composition are important, as Saturn's obliquity is slightly larger than the Earth's. Their model consists of a sum of 1D-photochemical model runs at different solar declinations and conditions. It does not include meridional transport processes nor the calculation of the actinic fluxes in 2D/3D. Saturn's high obliquity similarly impacts the stratospheric temperatures \citep{Fletcher2010}. This effect was accounted for by \citet{Moses2005b} in their photochemical model as part of a sensitivity case study, by locally warming their nominal temperature profile at two latitudes, according to the observations of \citet{Greathouse2005}. In this sense, the photochemical model in \citet{Guerlet2010} represents an improvement from the previous model of \citet{Moses2005b} as it includes the latitudinal thermal gradient observed both by \citet{Fletcher2007} and \citet{Guerlet2009}, but held constant with seasons. Finally, \citet{Moses2007} accounted for the meridional transport in a 2D-photochemical model, but similarly neglected the seasonal evolution of the stratospheric temperature. They were unable to reproduce the ground-based hydrocarbon observations prior to Cassini mission \citep{Greathouse2005}. After 10 years of Cassini measurements, data has shown that Saturn's stratospheric thermal structure is complex, with a 40K pole-to-pole gradient after solstice \citep{Fletcher2010}, and thermal oscillations in the equatorial zone \citep{Orton2008,Fouchet2008,Guerlet2011}.

For the moment, there is no 2D photochemical model that simultaneously accounts for seasonal forcing, meridional transport and the evolution of the stratospheric temperature. In this paper, we present a new step toward this model, applied to Saturn. These latitudinally and seasonally variable 1D models, coupled by a 3D-radiative transfer model, can be seen as an intermediate class of model between the 1D photochemical models that have the most complete chemistries and the GCMs that are focused on 3D dynamics. In this paper, we present a restricted version of our full-2D model. The goal of this preliminary study is to evaluate the atmospheric chemical response to seasonal forcing in terms of solar radiation and atmospheric temperatures. The meridional transport is therefore set to zero for this study in order to focus on photochemical effects. In forthcoming papers we will focus on the effect of 2D advective and diffusive transport on the predicted abundances.

In the first part of this paper, we present in detail how the seasonally variable parameters are accounted for in the model, including Saturn's orbital parameters and the thermal field. Then, we describe the photochemical model, the chemical scheme used in that model and the 3D radiative transfer model used to calculate the attenuation of the UV radiation in the atmosphere. We afterwards describe the seasonal evolution of the chemical composition, first by assuming that the thermal field does not evolve with time and latitude, to compare with previous findings, then by considering a more realistic thermal field with spatio-temporal variations. We underline the effect of such thermal field variations on the chemical composition. Finally, we will compare our results with the Cassini/CIRS observations.

\section{Seasonal modeling of the photochemistry}

\subsection{Introduction}

The amount of solar radiation striking the top of the atmosphere at a given latitude varies with seasons because of Saturn's obliquity and eccentricity. Atmospheric heating occurs through methane near-IR absorption of this radiation. Cooling is preponderant in the mid-IR range, mainly through emissions from acetylene, ethane, and, to a lesser extent, methane \citep{Yelle2001}. These IR-emissions increase with increasing atmospheric temperatures and/or abundances of these compounds. Therefore, the temperature field, as a function of altitude and latitude, mostly depends on the seasonal distribution of these species and on their response to the seasonally varying insolation.

Methane, which is generally assumed to be well-mixed in Saturn's atmosphere (see e.g., \citealt{Fletcher2009}) and optically thick in its IR bands, can be used as a thermometer to constrain the thermal field \citep{Greathouse2005}. Asymmetries in Saturn's atmospheric temperatures have been observed as a function of season, from Voyager \citep{Pirraglia1981, Hanel1981, Hanel1982, Conrath1983, Courtin1984} and ground-based observations (e.g., \citealt{Gillett1975, Rieke1975, Tokunaga1978, Gezari1989, Ollivier2000a, Greathouse2005}). These observations have been reproduced in an approximate sense by radiative transfer model predictions \citep{Cess1979, Bezard1984, Bezard1985}.

\begin{figure}[htp]
\begin{center}
\includegraphics[width=0.8\columnwidth]{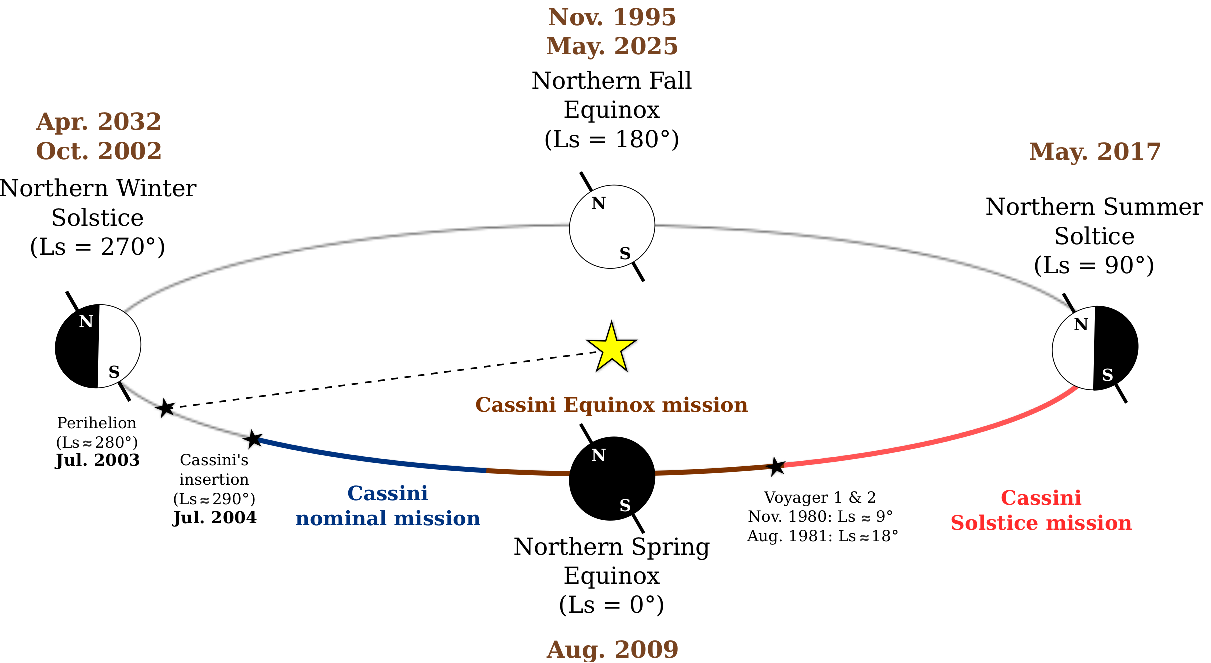}
\caption{Overview of Saturn's seasons. The position of Saturn on its orbit is defined by its heliocentric longitude ($L_s$). $L_s$ = 0$^{\circ}$, $L_s$ = 90$^{\circ}$, $L_s$ = 180$^{\circ}$ and $L_s$ = 270$^{\circ}$ correspond to Saturn's northern vernal equinox, summer solstice, autumnal equinox and winter solstice, respectively. The Cassini orbital insertion around Saturn occurred on July 1, 2004, shortly after the northern winter solstice (Oct. 2002) and Saturn's perihelion (Jul. 2003). Cassini's nominal, equinox and solstice missions are indicated. Voyager missions 1 and 2 flew by Saturn system on Nov. 12, 1980 and on Aug. 26, 1981, respectively, for their closest encounters.}
\label{fig:Seasons}
\end{center}
\end{figure}

The Cassini spacecraft arrived in Saturn's system in July 2004, shortly after its northern winter solstice  (see Fig. \ref{fig:Seasons}). It has provided full-coverage of the temperatures for the upper troposphere and stratosphere ever since. It has given us the opportunity to observe seasonal changes in the temperature field for over 10 years. For instance, the North/South thermal asymmetry at the northern winter solstice has been observed: the southern hemisphere was experiencing summer and was found warmer than the northern one \citep{Flasar2005, Howett2007, Fletcher2007}. Subsequently, Cassini observed how the winter hemisphere evolves when emerging from the shadow of the rings and how the summer hemisphere cools down when approaching equinox \citep{Fletcher2010, Sinclair2013}.

The main driver for atmospheric chemistry comes from solar UV radiation. This radiation initiates a complex chemistry through methane photolysis leading to the production of highly reactive chemical radicals. The kinetics of the chemical reactions triggered by photolysis generally have a thermal dependence that can impact the overall production/loss rates of atmospheric constituents over the course of Saturn's long seasons. 


Since we want to evaluate the atmospheric chemical response to seasonal forcing in terms of solar radiation and atmospheric temperatures, we thus compare the results of our model obtained in two different cases:
\begin{itemize}
   \item The temperature field consists of a single profile applied to all latitudes and seasons in a similar way to previous 1D studies. This study case will be denoted (U)
   \item The temperature field is vertically, latitudinally and seasonally variable. This study case will be denoted (S)
\end{itemize}
We stress again that the latitudes are not connected in the following study, i.e., the meridional transport is set to zero, so as to better quantify the effects of a seasonally variable temperature field on the distribution of chemical species. We defer the study of meridional transport to a forthcoming paper.

\subsection{Accounting for Saturn's eccentric orbit}
\label{subsection:Orbital}

\begin{figure}[htp]
\begin{center}
\includegraphics[width=10cm,keepaspectratio]{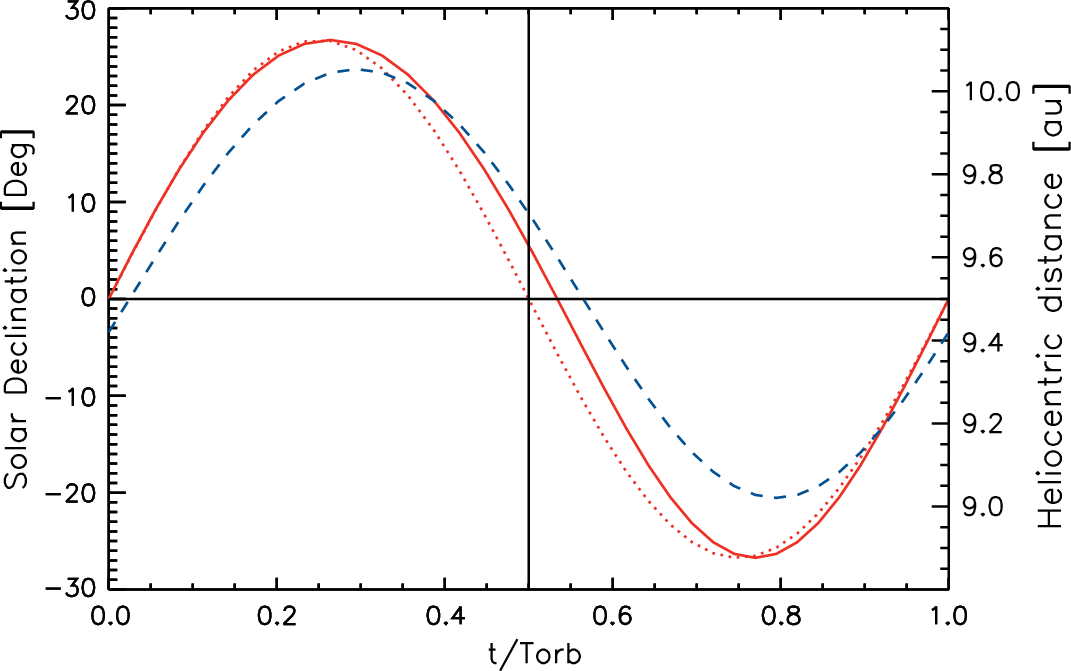}
\caption{Variation of solar declination (left scale) as a function of the orbital fraction, assuming Saturn's eccentricity (red solid line) and null eccentricity (red dotted line). The origin of the orbital fraction is taken at the northern spring equinox. The corresponding variation of the heliocentric distance (dashed line, right scale) is also plotted. Saturn's perihelion occurs shortly after the northern winter solstice. The solid vertical line at $t$/$T_{orb}$ $=$\,0.5 denotes the moment when the planet has spent half of its orbital period. At this point the subsolar point is still on the northern hemisphere, due to Saturn eccentricity.}
\label{fig:Declin}
\end{center}
\end{figure}

Due to Kepler's second law, Saturn's southern summer is shorter and hotter than the northern one, as Saturn reaches its perihelion shortly after the southern summer solstice (see Fig. \ref{fig:Declin}). In the present model, Saturn's elliptic orbit is sampled using a regularly spaced heliocentric longitude grid of 10$^{\circ}$. From one orbital point to the next one, the integration time of the photochemical model is computed from the Kepler equation and Saturn's true anomaly. The true anomaly and heliocentric longitude are similar quantities, only differing by their relative origin, the former one being Saturn's perihelion whereas the latter being the Vernal equinox. The offset position in heliocentric longitude of Saturn's perihelion was set at 280.077$^{\circ}$ \citep{Guerlet2014}, a value based on J2000 parameters. Similarly to \citet{Moses2005b}, integration over several orbits was needed for the simulations to converge down to the 100 mbar pressure level. Although the eddy diffusion coefficient profile as a function pressure was identical in every simulation, differences were found in the number of orbits required for convergence in the simulations depending on the thermal field. We will explain the reasons for these differences in section \ref{subsection:Seasonnal_Variability2}.

The variation of the solar declination as a function of the orbital fraction, starting from vernal equinox, is displayed in Fig. \ref{fig:Declin}. As Saturn's perihelion occurs shortly after the southern summer solstice, the orbital fraction during which the subsolar point is on the northern hemisphere is longer than the opposite one. This is shown by the solid curve being shifted to the right, at orbital fraction of 0.5, with respect to the circular case (dotted line).

\subsection{Temperature field}
\label{subsection:Temp_Prof}

\subsubsection{Spatially uniform thermal profile}
\label{subsubsection:Uniform}

In our first study case, the temperatures over the planet only vary with altitude and are constant with time and latitude, consistent with \citet{Moses2005b}. We have taken the thermal profile that was used to obtain the reduced chemical scheme \citep{Dob2011} we employ in our model. The temperatures below the $10^{-5}$\,mbar pressure level come from a retrieval performed by \citet{Fouchet2008} on Cassini/CIRS data observed at a planetographic latitude of 20$^{\circ}$S. Extrapolation to the upper stratosphere has been made using data from \cite{Smith1983} (see Fig. \ref{fig:profils_T}). This thermal profile is presented in Fig. \ref{fig:profils_T}. In what follows, we will refer to this case as the ``spatially uniform'' (U) temperature field case.

\subsubsection{Seasonally variable thermal field}
\label{subsubsection:Seasonal_T_map}
The second temperature field we considered comes from the seasonal radiative climate model of \citet{Greathouse2008}, which has already been compared to Cassini/CIRS observations \citep{Fletcher2010}. This radiative transfer model takes into account heating and cooling from Saturn's major atmospheric compounds, i.e. CH$_4$, C$_2$H$_2$, and C$_2$H$_6$, as well as seasonal variation of Saturn's orbital parameters, i.e., solar declination, heliocentric distance and eccentricity. It also includes ring shadowing and accounts for Saturn's oblate shape. In this second study case, the temperature varies with altitude, latitude and time. This case will be referred to as the "seasonal" (S) temperature field case.

\begin{figure}[htp]
\begin{center}
\includegraphics[width=0.4\columnwidth]{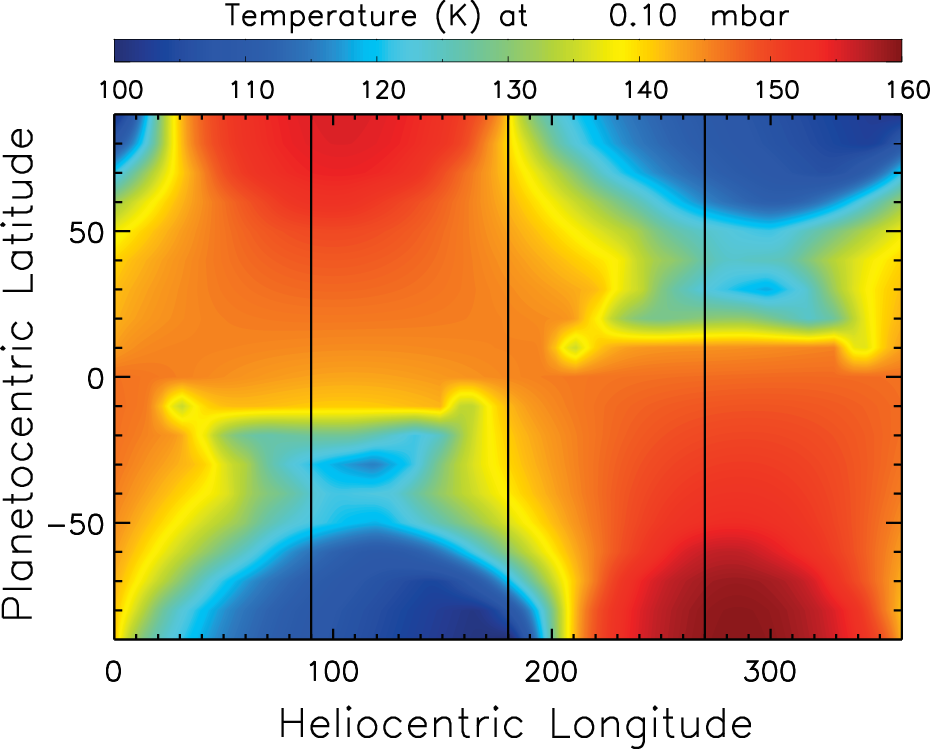}
\includegraphics[width=0.45\columnwidth]{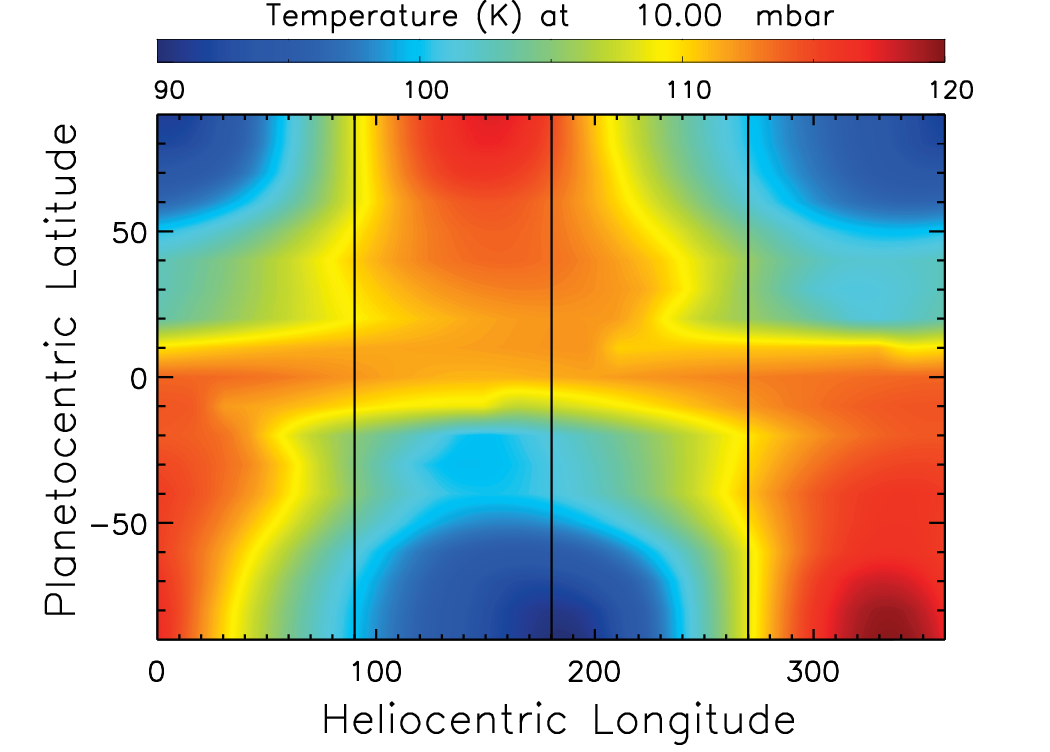}
\caption{Seasonal temperature field inferred from the radiative climate model of \citet{Greathouse2008} as a function of planetocentric latitude and heliocentric longitude. The temperature variation at 0.1 mbar is twice as large as that at 10 mbar due to the increase in thermal inertia with depth in the atmosphere (note the color range is stretched differently for the two plots). Left panel: Temperatures at 0.1\,mbar. Right panel: Temperatures at 10\,mbar.}
\label{fig:Temp_Greathouse}
\end{center}
\end{figure}

The seasonal thermal field used in this paper is shown in Fig. \ref{fig:Temp_Greathouse} as a function of planetocentric latitudes and heliocentric longitudes, and is presented for two pressure levels: 0.1\,mbar and 10\,mbar. Hereafter, all quoted latitudes are planetocentric, if not otherwise specified. The North-South asymmetry during the summer is caused by Saturn's eccentricity. The effects of ring shadowing are clearly observed at 0.1\,mbar, around the solstices between 0$^{\circ}$ and 40$^{\circ}$ planetocentric latitude in the winter hemispheres. Time-lag between temperatures and seasons, due to the atmospheric thermal-inertia, can be seen at 10\,mbar by the difference in temperature profile at 0.1 and 10\,mbar. The thermal field has been computed by taking, as a first guess, the CH$_4$, C$_2$H$_2$ and C$_2$H$_6$ vertical distributions observed by Cassini \citep{Guerlet2009} at planetographic latitude of 45$^{\circ}$S and held fixed with time.

\begin{figure}[htp]
\begin{center}
\includegraphics[width=1.\columnwidth]{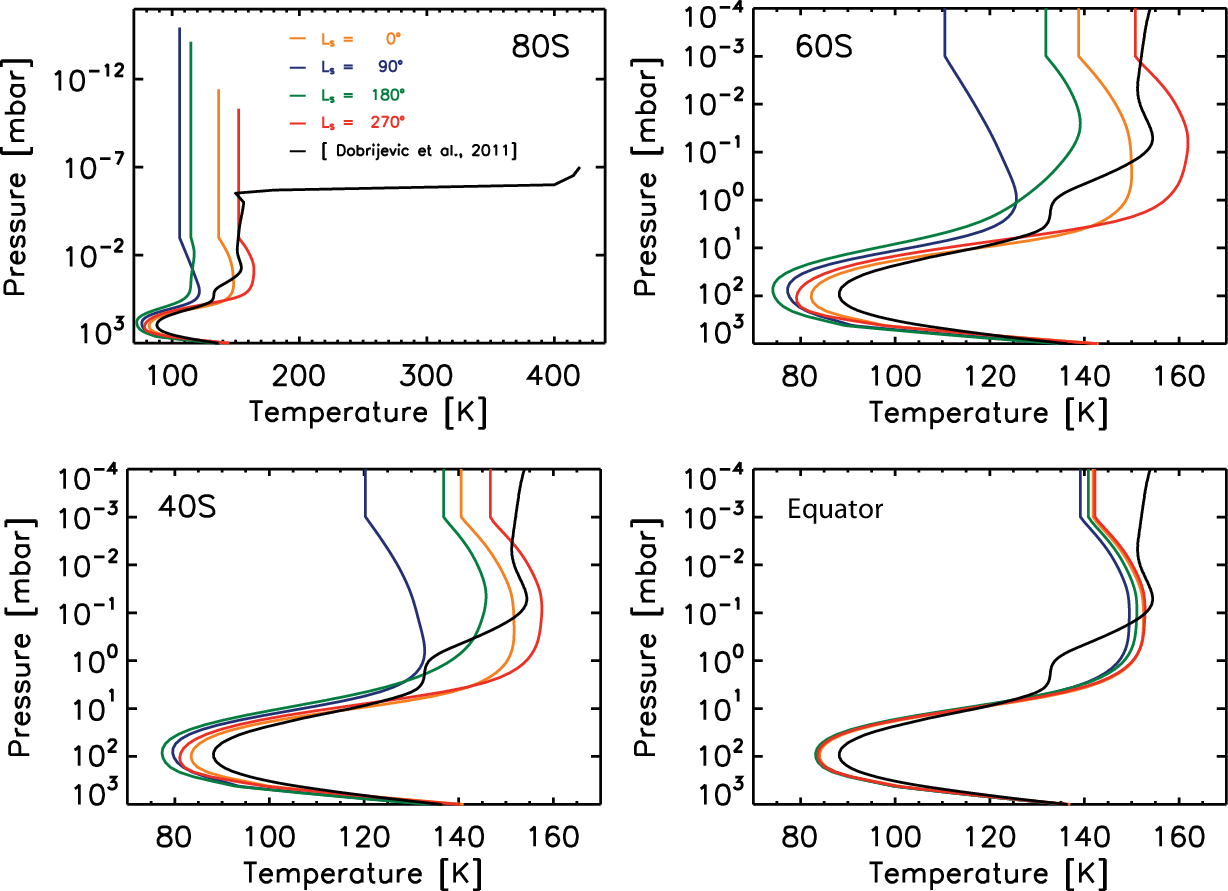}
\caption{Temperature profiles used in this work as a function of pressure. The colored lines depict the seasonally variable thermal field (S) predicted from the radiative climate model at the solstices and equinoxes, for 4 latitudes : 80$^{\circ}$S (upper-left panel), 60$^{\circ}$S (upper-right panel), 40$^{\circ}$S (lower left panel) and the equator (lower left panel). $L_s$ = 0$^{\circ}$, 90$^{\circ}$, 180$^{\circ}$ and 270$^{\circ}$ correspond to northern fall equinox, summer solstice, spring equinox and winter solstice, respectively (see Fig. \ref{fig:Seasons}). The black solid lines display the spatially uniform thermal field (U) we consider in this work. This profile comes from Cassini/CIRS observations \citep{Fouchet2008} and Voyager 2 observations \citep{Smith1983} (see text for details).
}
\label{fig:profils_T}
\end{center}
\end{figure}

The temperature map predicted from the radiative climate model is calculated within the pressure range from 500\,mbar to $10^{-6}$\,mbar \citep{Fletcher2010,Greathouse2008}. Although the seasonal model of \citet{Greathouse2008} extends down to 500\,mbar, temperatures are only accurate to 10\,mbar as this model was created primarily to model the stratosphere. At lower altitudes, the model lacks aerosol absorption and scattering and convective adjustement. Due to this lack of aerosols, the tropospheric temperatures are lower by 5-15 K than measured by Cassini. We note this discrepancy, but are focused on understanding effect of temperature on stratospheric photochemistry occur at altitudes above the 10 mbar level where the physics are self consistent.
Below the 500 mbar level we extrapolate the temperature assuming a dry adiabatic lapse rate, using a specific heat of $c_p$ = 10 658 J.kg$^{-1}$.K$^{-1}$ \citep{Irwin2006} and a latitude-dependent gravity field (see Supplementary Materials of \citealt{Guerlet2014}). Above $10^{-6}$\,mbar, where non-LTE effects dominate, the temperature was held constant, and no thermosphere is assumed above the stratosphere in this model. The lowest pressure level in our grid is set in order to ensure that each monochromatic optical depth is smaller than 1 in the UV at the top of the atmosphere. Fig. \ref{fig:profils_T} displays the resulting temperature profiles at 4 latitudes : 80$^{\circ}$S (upper-left panel), 60$^{\circ}$S (upper-right panel), 40$^{\circ}$S (lower left panel) and the equator (lower right panel). The colored solid lines represent the atmospheric temperatures inferred from the radiative climate model at solstices and equinoxes and the reconstruction procedure described above.

\subsubsection{Thermal evolution}
\label{subsubsection:Seasonal_Temp_Use}

The first case studied in this paper, namely the spatially uniform thermal field does not require special care on how the pressure-altitude background is treated, as it remains constant all along the year. However, when the temperature changes, i.e., in the case of a seasonally variable thermal field, the pressure-altitude background also changes and has to be handled carefully.

Two ways of dealing with changes in the atmospheric pressure-temperature background in photochemical modeling exist. Either the altitude grid is held constant and the pressure varies with temperature, or the pressure grid is held constant and the altitude grid is free to contract or expand (e.g., \citealt{Agundez2014}). Since the two approaches are self-consistent, we have choosen to hold the altitude grid constant and let the pressure grid vary with temperature.

This choice has been made to allow two latitudinally-contiguous numerical cells to exchange material through their common boundary, for future 2D-modeling including circulation and meridional advection. 


The altitude-temperature grid at all latitudes and heliocentric longitudes is built assuming hydrostatic equilibrium. Variations in scale height due to Saturn's latitudinally and altitudinally-dependent gravity field and variations in the mean molecular mass in the upper atmosphere due to molecular diffusion are included when solving the hydrostatic equilibrium equation.
The latitudinal-dependency adopted here follows the prescription of \citet{Guerlet2014}. We have made sure that the pressure-temperature background using this prescription is consistent with the latitudinally dependent gravitational field published by \citet{Lindal1985} and combined with the Voyager 2 zonal wind measurements \citep{Smith1982, Ingersoll1982}.

\begin{figure}[htp]
\begin{center}
\includegraphics[width=0.8\columnwidth]{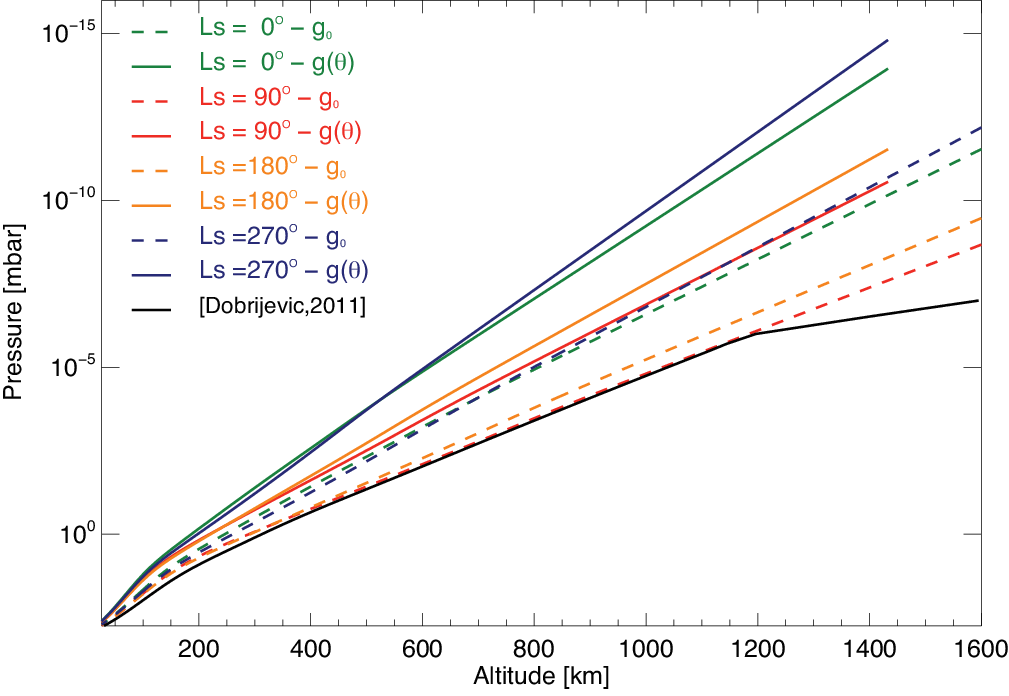}
\caption{Pressure-altitude grid at 80$^{\circ}$N for the solstices and equinoxes, assuming a constant surface gravity (dashed colored lines) and a latitudinally-variable surface gravity (solid colored lines). $L_s$ = 0$^{\circ}$, 90$^{\circ}$, 180$^{\circ}$ and 270$^{\circ}$ correspond to northern fall equinox, summer solstice, spring equinox and winter solstice, respectively (see Fig. \ref{fig:Seasons}). 0\,km is equal to the 1\,bar level. The black solid line represents the pressure-altitude grid of the uniform temperature profile.}
\label{fig:Construct}
\end{center}
\end{figure}

The effect on the pressure-altitude grid can be large as shown in Fig. \ref{fig:Construct}, which presents this grid for 80$^{\circ}$N at the equinoxes and solstices. Solid and dashed lines respectively represent this grid when the latitudinally-dependent gravity is included and when considering a constant surface gravity, set to the equatorial one. The pressure-altitude grid of the uniform model (see \ref{subsubsection:Uniform}) is also shown (black solid line) for comparison. Differential surface gravity due to Saturn's high rotation rate results in more contracted atmospheric columns at polar latitudes. Hence, at the same altitude level, the pressure is lower at the poles relative to the equator when considering variable surface gravity. At a given latitude, the column also expands or contracts with temperature as shown in Fig. \ref{fig:Construct}. This example at 80$^{\circ}$ N is extreme as the amplitude of the temperature variation with season is maximum at polar latitudes. The seasonal variation of the pressure-altitude grid is damped toward the equator, as the seasonal thermal gradient is reduced in this region.

In the present paper, we have chosen to work with a common altitude grid for all latitudes and seasons that start at a common origin ($z$\,$=$\,$0$\,km and $P$\,$=$\,$1$\,bar). Above that origin, the pressure grid expands or contracts according to temperature changes.
The mole fraction vertical profiles of the model species are expressed as a function of pressure and thus follow the same contraction/expansion as the pressure grid. Therefore, each time the temperature/pressure grid changes, the mole fraction profiles are interpolated onto the new pressure grid.

\begin{figure}[htp]
\begin{center}
\includegraphics[width=0.8\columnwidth]{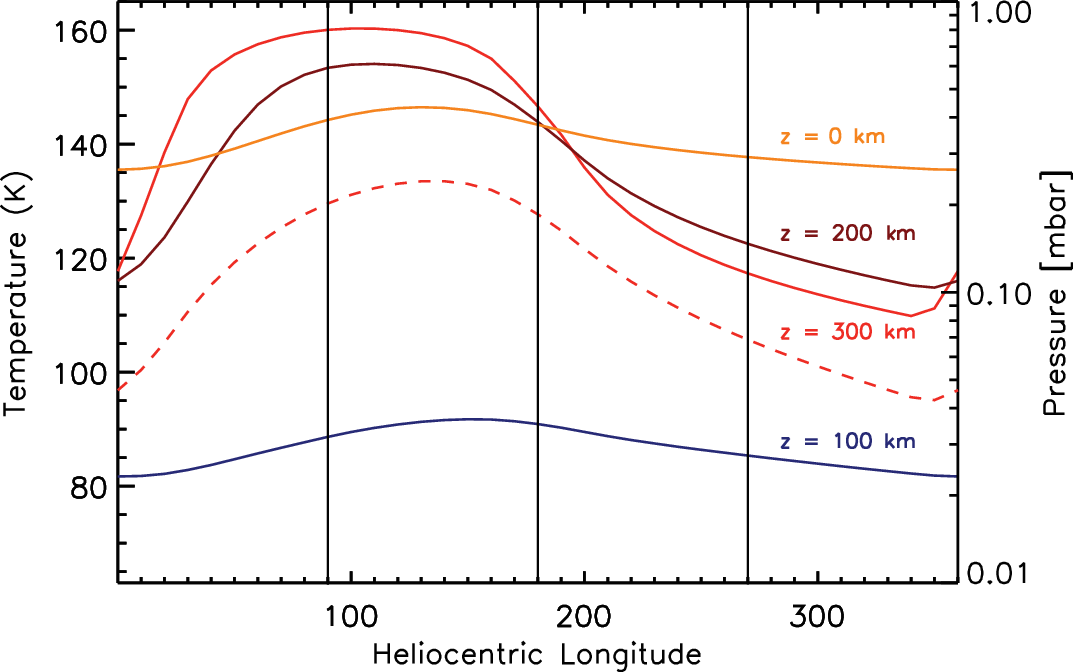}
\caption{Seasonal evolution of temperature (left scale, solid lines) and pressure (right scale, dashed line) at altitudes of 300\,km (red lines), 200\,km (brown), 100\,km (blue) and 0\,km (orange). The quantities are presented for a planetocentric latitude of 80$^{\circ}$N, where the variations are most noticeable. The black solid lines indicate the position of the solstices and equinoxes (see Fig. \ref{fig:Seasons}).}
\label{fig:Altitude_Plot_1}
\end{center}
\end{figure}

It is instructive to represent the seasonal evolution of the temperature and pressure at a given latitude for a few altitude levels (see Fig. \ref{fig:Altitude_Plot_1} for an illustration at 80$^{\circ}$N). When the temperature rises, the atmospheric column expands, and the associated pressure at the same altitude increases. It should be noted that temperature and pressure are not totally in phase, as the pressure at a given altitude depends on the thermodynamical conditions of the altitude levels underneath. Therefore the temperature at a few altitude levels below the considered altitude are presented on the same figure. We note that the increase in the pressure at 300 km is in phase with the temperature changes at altitude levels below that level.


\section{Latitudinally and seasonally variable 1D models}
\label{section:model}

\subsection{General description}
\label{subsection:Gen_Descrip}

In an atmosphere, the spatio-temporal distribution of each species' number density is governed by the continuity-transport equation, that is:
\begin{equation}
\dfrac{\partial n_i}{\partial t} = P_i - n_i L_i - \mathbf{ \nabla} \cdot ( \mathbf{\Phi_i}) \label{eq:continuity}
\end{equation}
where $n_i$ [cm$^{-3}$] is the number density, $P_i$ [cm$^{-3}$\,s$^{-1}$] the (photo)chemical production rate, $L_i$ [s$^{-1}$] the (photo)chemical loss rate and $\Phi_i$ [cm$^{-2}$\,s$^{-1}$] is the particle flux due to transport. Longitudinal mixing timescales appear to be relatively short in Jupiter's atmosphere (\textit{e.g.,} \citealt{Banfield1996}) and deviations from the mean zonal temperatures are limited \citep{Flasar2004}. We assume the situation is similar at Saturn and we thus do not consider longitudinal variability in this study. The continuity equation is then solved on a 2D altitude-latitude spherical grid. We use a 13\,km altitude grid resolution, in order to have at least 3 altitudinal numerical cells per scale height at all times throughout the year. The planet radius considered here is Saturn's mean radius, $R = 58,210$ km \citep{Guillot2005}, which corresponds to the altitude level $z$\,=\,0\,km. The flux $\Phi_i$ includes transport processes in the vertical and the meridional directions.

Taking these mixing processes into account at all scales and in detail would require a full hydrodynamical model, which is beyond the scope of this work. In our model, the physical processes that are accounted for through the vertical flux $\Phi^{z}_i$, are eddy diffusion, molecular diffusion and vertical advection. This flux is expressed as:

\begin{align}
\Phi^{z}_i =& - D_i n_i \left( \frac{1}{y_i} \dfrac{\partial y_i}{\partial z} + \frac{1}{H_i} - \frac{1}{H} \right) - K_{zz} n_i \left( \frac{1}{y_i} \dfrac{\partial y_i}{\partial z} \right)  + v_i^z n_i \label{eq:flux_z}
\end{align}
where $y_i$ is the mole fraction of species $i$, defined as the ratio between the number density of $i$ over the total number density. $H_i$ and $H$ [cm] are respectively the specific and the mean density scale height, $D_i$ [cm$^{2}$\,s$^{-1}$] the molecular diffusion coefficient, $K_{zz}$ [cm$^{2}$\,s$^{-1}$] the vertical eddy diffusion coefficient and $v_i^z$ [cm\,s$^{-1}$] the vertical wind. The numerical scheme used in this study is similar to the one used by \citet{Agundez2014} in their pseudo-2D photochemical model except that we use an upwind scheme to treat the advective part of the molecular diffusion \citep{Godunov1959}. The meridional flux $\Phi^{\theta}_i$ is set to zero for the current study.


The vertical eddy diffusion coefficient $K_{zz}$ is a free parameter in the model to account for mixing processes caused by dynamics occuring at every scale. This parameter may vary with altitude and latitude, but our knowledge for giant planet stratospheres is very limited (see for instance \citealt{Moreno2003} and \citealt{Liang2005}). This coefficient is related to the small-scale waves and is therefore expected to be influenced by the atmospheric number density \citep{Lindzen1971,Lindzen1981}. Consequently, 2D/3D models will probably have to account for its latitudinal and longitudinal variability. In this study, we consider that $K_{zz}$ is fixed with respect to the pressure coordinate. Due to the lack of constraint on that parameter, we consider this does not vary with latitude.
The reduced chemical scheme we use has been obtained using the $K_{zz}$ profile of \citet{Dob2011}. Therefore, we consistently take their $K_{zz}$. 
The molecular diffusion coefficient we adopt is based on experimental measurements of binary gas diffusion coefficients \citep{Fuller1,Fuller2}. As a first step in this study, we set $K_{yy}$, $v^z$ and $v^{\theta}$ to zero. These parameters will be studied in a forthcoming paper, either by trying to fit them from the observations or by testing outputs of the yet-to-be-finalized GCM of \citet{Guerlet2014}.


At the lower boundary of the model (i.e. 1\,bar), the H$_2$ and He mole fractions are set to 0.8773 and 0.118, respectively \citep{Conrath2000}. The methane mole fraction was set to 4.7$\times$ $10^{-3}$ according to recent Cassini/CIRS observations \citep{Fletcher2009}. At this boundary, all other compounds diffuse down to the lower troposphere at their maximum diffusion velocity, i.e., $v$ =  -$K_{zz}(0) / H(0)$. At the upper boundary of the model, all fluxes are set to zero except for atomic hydrogen. Following \citet{Moses2005a}, we set its influx to $\Phi_H = 1.0 \times 10^8$ cm$^{-2}$\,s$^{-1}$ at all latitudes.

\subsection{Chemical Scheme}
\label{subsection:Network}

In typical 1D photochemical models, the chemical schemes contain as many reactions as possible, i.e., usually hundreds, and numerous species. This makes it extremely difficult for current computers to solve equation (\ref{eq:continuity}) in a reasonable time when extending such models to 2D or 3D. \citet{Dob2011} have developed an objective methodology to reproduce the chemical processes for a subset of compounds of interest (usually observed compounds) with a limited number of reactions. These are extracted from a more complete chemical scheme by running a 1D photochemical model and applying propagation of uncertainties on chemical rates and a global sensitivity analysis.

Uncertainties in the chemical rate constants are a critical source of uncertainty in photochemical model predictions \citep{Dobrijevic1998}, as chemical schemes generally include tens to hundreds of chemical compounds, non-linearly coupled in even more reactions. Propagating uncertainties on each chemical reaction, using a Monte Carlo procedure for instance, can lead to several orders of magnitude in uncertainty \citep{Dobrijevic2003}. By computing correlations between reaction rate uncertainties and photochemical model predictions, \citet{Dobrijevic2010a} and \citet{Dobrijevic2010b} developed a global sensitivity analysis methodology to identify key reactions in chemical schemes. These reactions have a major impact on the results, either because their uncertainty is intrinsically high, or because they significantly contribute in the production/loss terms of the compound of interest (or one of the compounds related to it). 

A reaction that has a low degree of significance means that changing its rate constant (within its uncertainty range) does not significantly change the results of the model, or well inside the model error bars. Building a reduced network then consists in removing reactions, and thus compounds once they are no longer linked by reactions, that have a very low degree of significance. The results stay very close to the median profile of the full chemical scheme for the remaining compounds.

\begin{figure}[htp]
\begin{center}
\includegraphics[width=0.8\columnwidth]{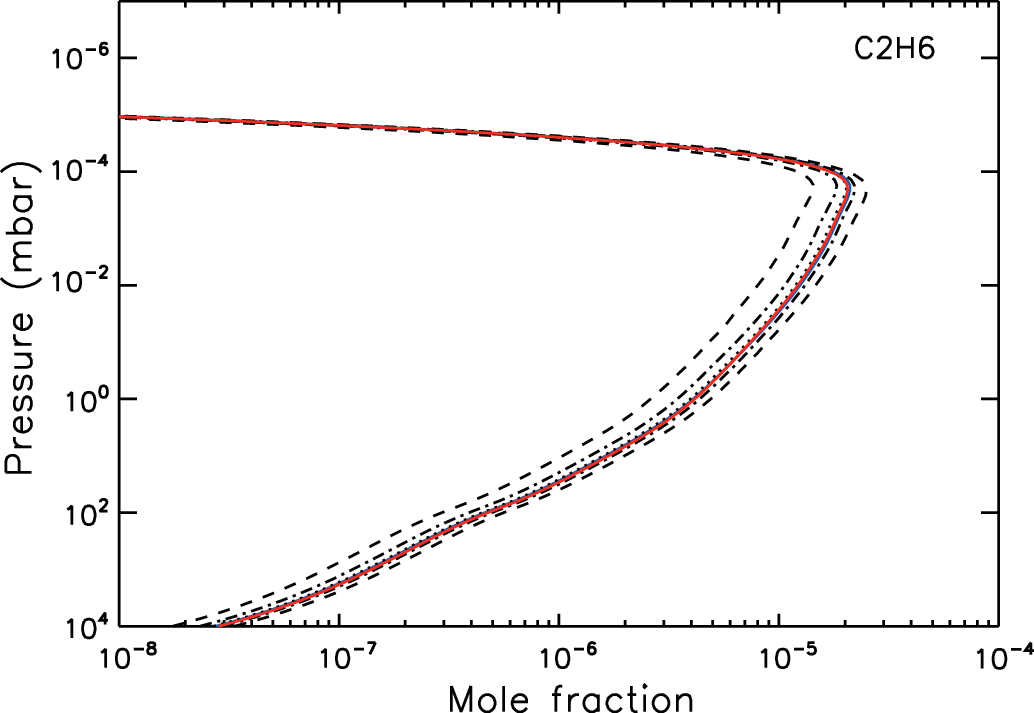}\\
\includegraphics[width=0.8\columnwidth]{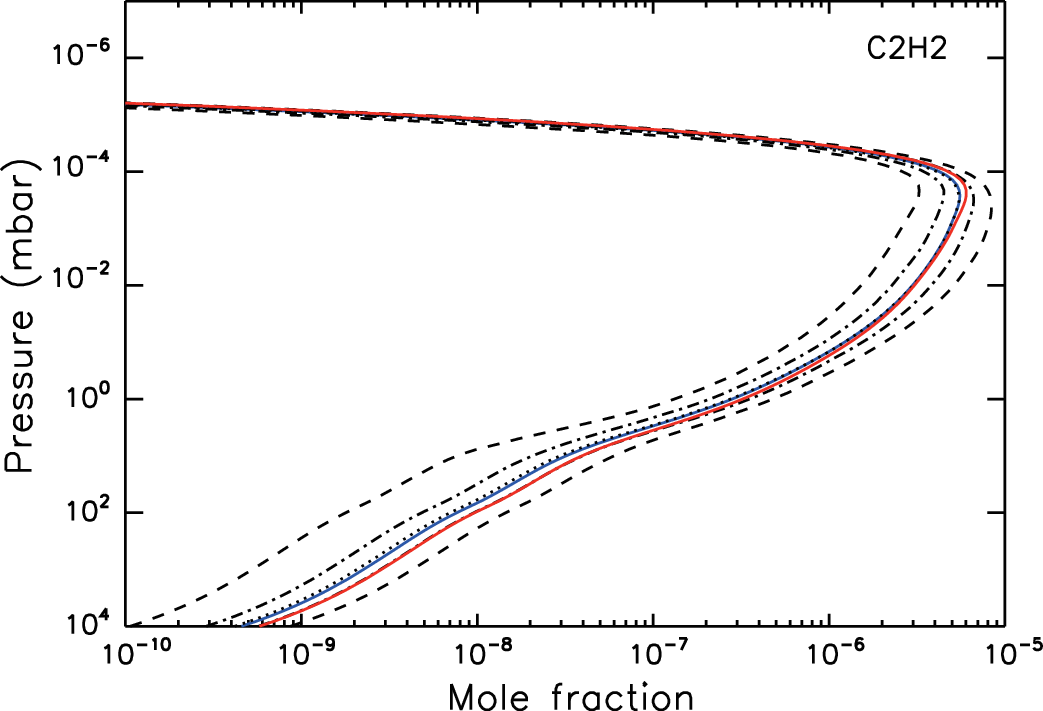}
\caption{Red solid line: C$_2$H$_6$ (top panel) and C$_2$H$_2$ (botom panel) vertical profile with the reduced chemical scheme. Blue line: nominal vertical profile obtained using the initial chemical scheme. Black dotted line: median profile of the full-scheme distribution. Black dashed-dotted lines: 5th and 15th 20-quantiles of the full-scheme distribution. Black dashed lines: 1th and 19th 20-quantiles of the full-scheme distribution.}
\label{fig:Reduction}
\end{center}
\end{figure}

A reduced chemical scheme is valid when it agrees with the full chemical scheme, given the uncertainties of each chemical compound profile. The initial scheme we consider includes 124 compounds, 1141 reactions and 172 photodissociation processes and comes from \citet{Loison2014}. The compounds we have selected to build the reduced chemical scheme are the ones monitored by Cassini/CIRS and most relevant regarding stratospheric heating and cooling: CH$_4$, C$_2$H$_2$, and C$_2$H$_6$  \citep{Guerlet2009,Sinclair2013}. We based our reduction scheme on the model validation performed for Saturn's hydrocarbons by \citealt{Cavalie2015}. C$_2$H$_2$, and C$_2$H$_6$ vertical profiles using the reduced chemical scheme are in good agreement with the full chemical scheme results (Fig. \ref{fig:Reduction}). The reduced scheme produces vertical profiles that are within the 5th and 15th of the full-scheme 20-quantiles distribution for C$_2$H$_6$ at all pressure levels and for C$_2$H$_2$ above 10\,mbar. Below 10\,mbar, the C$_2$H$_2$ vertical profile is almost superimposed to the 15th 20-quantiles of the distribution.

Three main oxygen compounds have also been added to the reduced scheme which are present in Saturn's stratosphere (H$_2$O, CO, and CO$_2$) as ground work for a forthcoming paper on the spatial distribution of H$_2$O, following observations by Herschel \citep{Hartogh2009,Hartogh2011}. The oxygen species will not be used in the present study and will not be presented nor discussed any further. In the end, the reduced scheme used in the present study includes 22 compounds, 33 reactions, and 24 photodissociations, listed in Table \ref{tab:Chem_Network}. Such a reduced chemical scheme enables extending photochemical computations to 2D/3D.




\begin{table}
\begin{center}
\caption{List of the 22 chemical compounds included in the scheme }
\label{tab:Chem_Network}
\begin{tabular}{l}
\hline
He; H; H$_2$ \\
CH; C; $^{1}$CH$_2$; $^{3}$CH$_2$; CH$_3$; CH$_4$ \\
C$_2$H; C$_2$H$_2$; C$_2$H$_3$; C$_2$H$_4$; C$_2$H$_5$; C$_2$H$_6$ \\
O$^3$P; O$^1$D; OH; H$_2$O \\
CO; CO$_2$; H$_2$CO \\
\hline
\end{tabular}
\end{center}
\end{table}


\LTcapwidth=\textwidth

\begin{center}
\begin{longtable}{llll}\caption{List of reactions of the reduced network (references can be found in \citet{Loison2014}). $k(T)$ = $\alpha \times (T/300)^{\beta} \times \textrm{exp}(-\gamma/T)$ in cm$^3$ molecule$^{-1}$ s$^{-1}$ or cm$^6$ molecule$^{-1}$ s$^{-1}$. $k_{adduct}$ = ($k_0$ [M] $F$ + $k_r$) $k_{\infty}$/$k_0$[M] + $k_{\infty}$ with log($F$) = log($F_c$)/1 + [log($k_0$[M] + $k_{capture}$)/$N$]$^2$, $F_c$ = 0.60 and $N$ = 1. Please refer to \citet{Hebrard2013} for details about the semi-empirical model.} 

\label{tab:Chem_Reactions} \\

\hline
 & Reactions & & Rate coefficients \\
\hline

R1 & H + CH & $\rightarrow$ C + H$_2$  &    $1.24 \times 10^{-10} \times (T/300)^{0.26}$      \\

R2 & H + $^3$CH$_2$ & $\rightarrow$ CH + H$_2$  &    $2.2 \times 10^{-10} \times (T/300)^{0.32}$               \\

R3 & H + $^3$CH$_2$ & $\rightarrow$  CH$_3$  &   $k_0 = 3.1 \times 10^{-30} \times \textrm{exp}(457/T) $              \\
& &   &    $k_{\infty} = 1.5 \times 10^{-10} $            \\
& &   &    $k_r = 0 $            \\

R4 & H + CH$_3$ & $\rightarrow$  CH$_4$  &   $k_0 = 8.9 \times 10^{-29}  \times (T/300)^{-1.8} \times \textrm{exp}(-31.8/T) $              \\
& &   &    $k_{\infty} = 3.2 \times 10^{-10}  \times (T/300)^{0.133} \times \textrm{exp}(-2.54/T) $             \\
& &   &    $k_r = 1.31 \times 10^{-16}  \times (T/300)^{-1.29} \times \textrm{exp}(19.6/T) $            \\

R5 &H + C$_2$H$_2$ & $\rightarrow$  C$_2$H$_3$  &  $k_0 = 2.0 \times 10^{-30}  \times (T/300)^{-1.07} \times \textrm{exp}(-83.8/T) $              \\
& &   &    $k_{\infty} = 1.17 \times 10^{-13}  \times (T/300)^{8.41} \times \textrm{exp}(-359/T) $             \\
& &   &    $k_r = 0 $            \\

R6 &H + C$_2$H$_3$ & $\rightarrow$ C$_2$H$_2$ + H$_2$  &   $6.0 \times 10^{-11} $           \\

R7 & H + C$_2$H$_3$ & $\rightarrow$  C$_2$H$_4$  &  $k_0 = 3.47 \times 10^{-27}  \times (T/300)^{-1.3}  $              \\
& &   &    $k_{\infty} = 1.0 \times 10^{-10} $             \\
& &   &    $k_r = 0 $            \\

R8 &H + C$_2$H$_4$ & $\rightarrow$  C$_2$H$_5$  &    $k_0 = 1.0 \times 10^{-29}  \times (T/300)^{-1.51} \times \textrm{exp}(-72.9/T) $              \\
& &   &    $k_{\infty} = 6.07 \times 10^{-13}  \times (T/300)^{-5.31} \times \textrm{exp}(174/T) $             \\
& &   &    $k_r = 0 $            \\

R9 & H + C$_2$H$_5$ & $\rightarrow$ C$_2$H$_6$  &     $k_0 = 2.0 \times 10^{-28}  \times (T/300)^{-1.5}  $             \\
& &   &    $k_{\infty} = 1.07 \times 10^{-10} $             \\
& &   &    $k_r = 0 $            \\

R10 & H + C$_2$H$_5$ & $\rightarrow$ CH$_3$ + CH$_3$  &  $k_0 = k_{\infty} - k_{adduct} $              \\

R11 &C + H$_2$ & $\rightarrow$  $^3$CH$_2$  &     $k_0 = 7.0 \times 10^{-32}  \times (T/300)^{-1.5} $             \\
& &   &    $k_{\infty} = 2.06 \times 10^{-11}  \times \textrm{exp}(-57/T) $             \\
& &   &    $k_r = 0 $            \\

R12 &CH + H$_2$ & $\rightarrow$ CH$_3$  &   $k_0 = 6.2 \times 10^{-30}  \times (T/300)^{-1.6} $               \\
& &   &    $k_{\infty} = 1.6 \times 10^{-10}  \times (T/300)^{-0.08}  $             \\
& &   &    $k_r = 0 $            \\

R13 &CH + CH$_4$ & $\rightarrow$ C$_2$H$_4$ + H  &       $1.05 \times 10^{-10} \times (T/300)^{-1.04} \times \textrm{exp}(-36.1/T)$        \\

R14 & $^1$CH$_2$ + H$_2$ & $\rightarrow$ $^3$CH$_2$ + H$_2$  &    $1.6 \times 10^{-11} \times (T/300)^{-0.9} $            \\

R15 & $^1$CH$_2$ + H$_2$ & $\rightarrow$ CH$_3$ + H  &     $8.8 \times 10^{-11} \times (T/300)^{0.35} $            \\

R16 & $^3$CH$_2$ + H$_2$ & $\rightarrow$ CH$_3$ + H  &     $8.0 \times 10^{-12} \times \textrm{exp}(-4500/T) $             \\

R17 & $^3$CH$_2$ + CH$_3$ & $\rightarrow$ C$_2$H$_4$ + H  &   $1.0 \times 10^{-10} $              \\

R18 & $^3$CH$_2$ + C$_2$H$_3$ & $\rightarrow$ C$_2$H$_2$ + CH$_3$  &  $3.0 \times 10^{-11} $               \\

R19 & $^3$CH$_2$ + C$_2$H$_5$ & $\rightarrow$ C$_2$H$_4$ + CH$_3$  &  $3.0 \times 10^{-11} $               \\

R20 & CH$_3$ + CH$_3$ & $\rightarrow$ C$_2$H$_6$  &  $k_0 = 1.8 \times 10^{-26}  \times (T/300)^{-3.77}  \times \textrm{exp}(-61.6/T) $               \\
& &   &    $k_{\infty} = 6.8 \times 10^{-11}  \times (T/300)^{-0.359}  \times \textrm{exp}(-30.2/T) $             \\
& &   &    $k_r = 0 $            \\

R21 & C$_2$H + H$_2$ & $\rightarrow$ C$_2$H$_2$ +  H  &   $1.2 \times 10^{-11} \times \textrm{exp}(-998/T) $              \\

R22 & C$_2$H + CH$_4$ & $\rightarrow$ C$_2$H$_2$ +  CH$_3$  &   $1.2 \times 10^{-11} \times \textrm{exp}(-491/T) $               \\

R23 & C$_2$H$_3$ + H$_2$ & $\rightarrow$ C$_2$H$_4$ +  H  &     $3.45 \times 10^{-14} \times (T/300)^{2.56} \times \textrm{exp}(-2530/T) $             \\

R24 & C$_2$H$_3$ + CH$_4$ & $\rightarrow$ C$_2$H$_4$ +  CH$_3$  &   $2.13 \times 10^{-14} \times (T/300)^{4.02} \times \textrm{exp}(-2750/T) $                 \\

R25 & O($^3$P) + CH$_3$ & $\rightarrow$ CO +  H$_2$ + H  &   $2.9 \times 10^{-11} $              \\

R26 & O($^3$P) + CH$_3$ & $\rightarrow$ H$_2$CO + H  &    $1.1 \times 10^{-10} $             \\

R27 & O($^3$P) + C$_2$H$_5$ & $\rightarrow$ OH + C$_2$H$_4$  &   $3.0 \times 10^{-11} $             \\

R28 & O($^1$D) + H$_2$ & $\rightarrow$ OH + H  &   $1.1 \times 10^{-10} $              \\

R29 & OH + H$_2$ & $\rightarrow$ H$_2$O + H  &    $2.8 \times 10^{-12} \times \textrm{exp}(-1800/T) $              \\

R30 & OH + CH$_3$ & $\rightarrow$ H$_2$O + $^1$CH$_2$  &   $3.2 \times 10^{-11} $              \\

R31 & OH + CH$_3$ & $\rightarrow$ H$_2$CO + H$_2$  &    $8.0 \times 10^{-12} $             \\

R32 & OH + CO & $\rightarrow$ CO$_2$ + H  &    $1.3 \times 10^{-13} $             \\

R33 & H$_2$CO + C & $\rightarrow$ CO + $^3$CH$_2$  &    $4.0 \times 10^{-10} $             \\

\hline
\end{longtable}
\end{center}

\medskip

\begin{table}[h!]
\begin{center}
\caption{Photodissocation processes (References can be found in \citealt{Loison2014})}
\label{tab:Photolysis}
\begin{tabular}{lll}
\hline
\hline
& Photodissociations  \\
\hline

R34 & OH + $h \nu$   & $\rightarrow$  O($^1$D) + H  \\

R35 & H$_2$O + $h \nu$   & $\rightarrow$   H + OH \\
R36 & 			       & $\rightarrow$   H$_2$ +  O($^1$D) \\
R37 &  			       & $\rightarrow$   H + H + O($^3$P) \\

R38 & CO + $h \nu$   & $\rightarrow$   C + O($^3$P) \\

R39 & CO$_2$ + $h \nu$   & $\rightarrow$   C + O($^1$D) \\
R40 & 			    & $\rightarrow$   CO + O($^3$P) \\

R41 & H$_2$ + $h \nu$   & $\rightarrow$   H + H \\

R42 & CH$_4$ + $h \nu$   & $\rightarrow$   CH$_3$ + H \\
R43 &			       & $\rightarrow$   $^1$CH$_2$ + H + H  \\
R44 &			       & $\rightarrow$   $^1$CH$_2$ + H2  \\
R45 &			       & $\rightarrow$   $^3$CH$_2$ + H + H  \\
R46 &			       & $\rightarrow$   CH + H$_2$ + H  \\

R47 & CH$_3$ + $h \nu$   & $\rightarrow$   $^1$CH$_2$ + H \\

R48 & C$_2$H$_2$ + $h \nu$   & $\rightarrow$   C$_2$H + H \\

R49 & C$_2$H$_3$ + $h \nu$   & $\rightarrow$   C$_2$H$_2$ + H \\

R50 & C$_2$H$_4$ + $h \nu$   & $\rightarrow$   C$_2$H$_2$ + H$_2$ \\
R51 & 					   & $\rightarrow$   C$_2$H$_2$ + H + H \\
R52 & 					   & $\rightarrow$   C$_2$H$_3$ + H \\

R53 & C$_2$H$_6$ + $h \nu$   & $\rightarrow$   C$_2$H$_4$ + H$_2$ \\
R54 &    	  			       & $\rightarrow$   C$_2$H$_4$ + H + H \\
R55 &    	  			       & $\rightarrow$   C$_2$H$_2$ + H$_2$ + H$_2$ \\
R56 &    	  			       & $\rightarrow$   CH$_4$ + H$_2$ + H$_2$ \\
R57 &    	  			       & $\rightarrow$   CH$_3$ + CH$_3$ \\

\hline
\hline
\end{tabular}

\end{center}
\end{table}

\subsection{Actinic Flux}
\label{subsection:Radiative_Tranfer}

The knowledge of the solar UV flux at any latitude/altitude/season is required to properly compute photodissociation coefficients. We use a full-3D spherical line-by-line radiative transfer model, improved over the model initially developed by \citet{Brillet1996}, to account for the attenuation of solar UV in the atmosphere. It now accounts for the full 3D distribution of absorbers instead of assuming vertically homogeneous distributions in latitude and longitude as in \citet{Brillet1996}. However, as stated previously, we consider here a zonally mixed atmosphere and limit variability to altitude and latitude. Absorption is formally calculated by the exact computation of the optical path. Rayleigh diffusion is also accounted for using single photon ray tracing in a Monte Carlo procedure. The wavelengths considered here range from 10 nm to 250 nm, because the hydrocarbons considered in this study do not substantially absorb beyond these limits. The radiative transfer procedure uses the altitude-latitude absorption and diffusion coefficients, extrapolated onto a 3D atmosphere assuming zonal homogeneity. Correspondence between subsolar and planetocentric coordinates is then made assuming Saturn's orbital parameters at the moment of the Kronian year, i.e., when the altitude-latitude-longitude actinic flux needs to be computed. Saturn's ellipsoidal shape is not taken into account, while the elliptical orbit is. The elliptical orbit causes a peak in actinic flux during southern summer.

\begin{figure}[htp]
\begin{center}
\includegraphics[width=0.8\columnwidth]{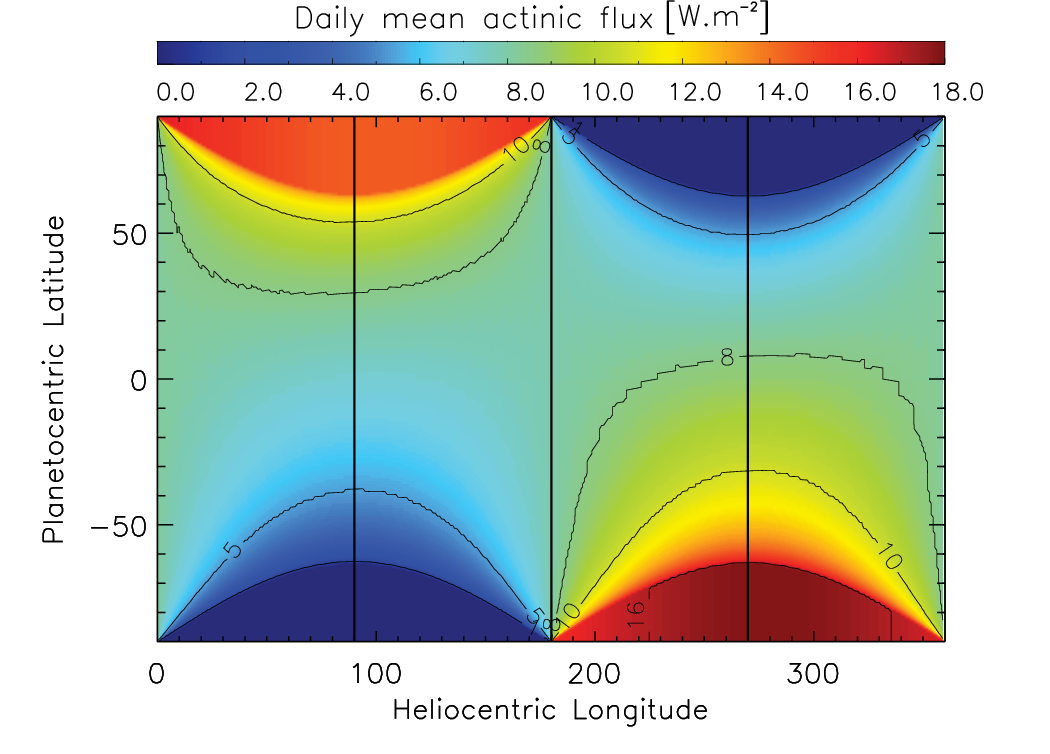}
\includegraphics[width=0.85\columnwidth]{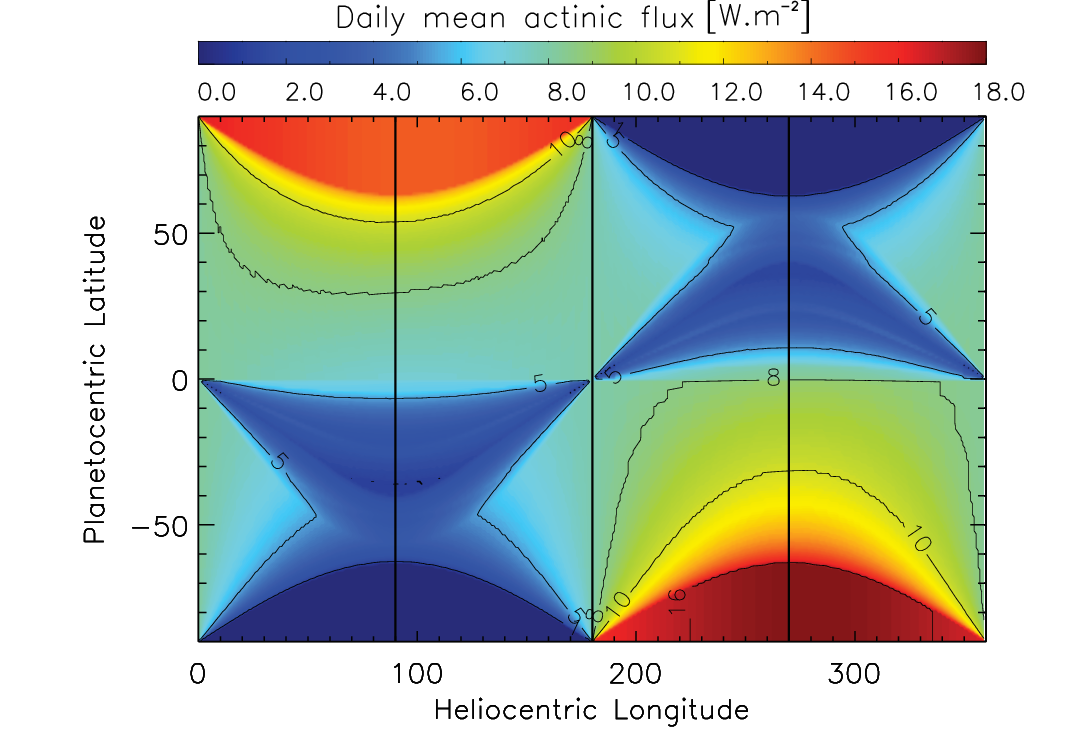}
\caption{Daily mean actinic flux in [W\,m$^{-2}$] as a function of planetocentric latitude and heliocentric longitude. Ring shadowing is included in the lower panel.  The black solid lines indicate the position of the solstices and equinoxes (see Fig. \ref{fig:Seasons}).}
\label{fig:Actinic}
\end{center}
\end{figure}

\begin{figure}[htp]
\begin{center}
\includegraphics[width=0.8\columnwidth]{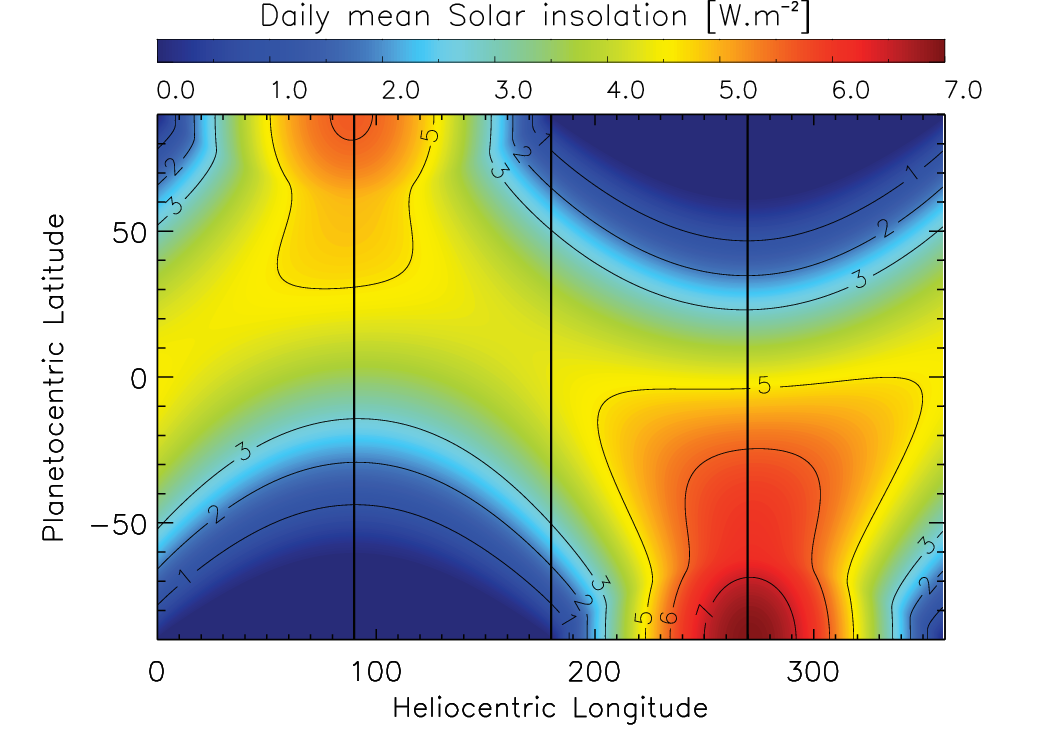}
\includegraphics[width=0.8\columnwidth]{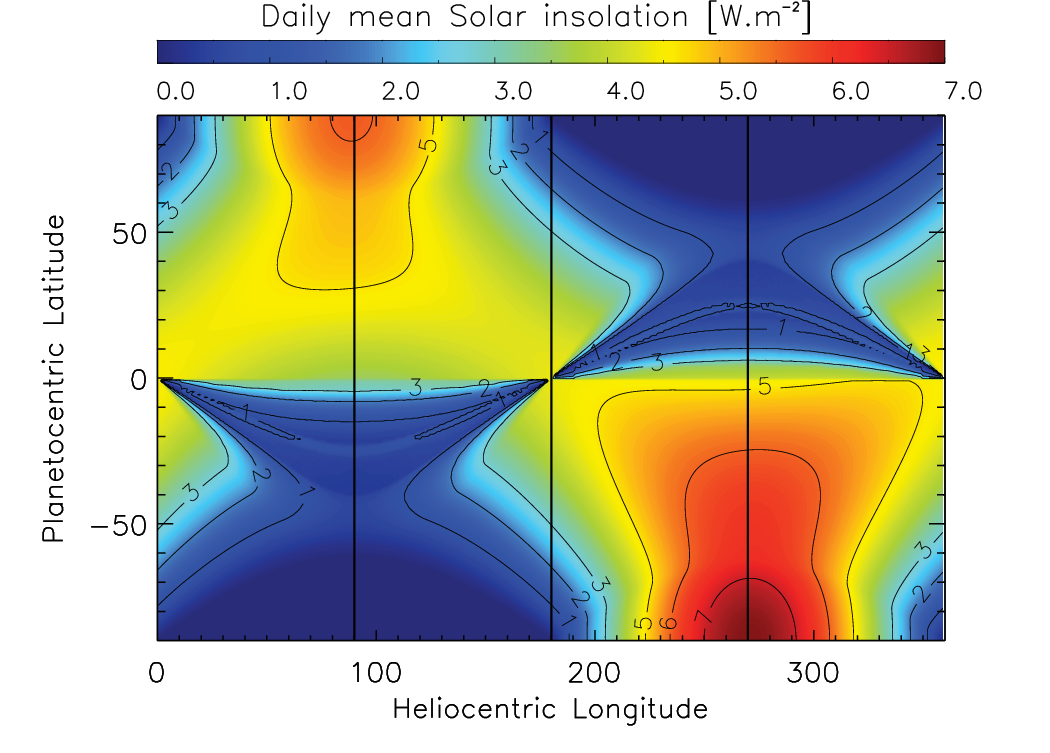}
\caption{Daily mean insolation in [W\,m$^{-2}$] as a function of planetocentric latitude and heliocentric longitude (i.e., seasons) received by a horizontal unit surface in Saturn's atmosphere. Ring shadowing is included in the lower panel. The black solid lines indicate the position of the solstices and equinoxes (see Fig. \ref{fig:Seasons}).}
\label{fig:Insolation}
\end{center}
\end{figure}

The daily-averaged actinic flux [W\,m$^{-2}$] at the top of the atmosphere is shown in Fig. \ref{fig:Actinic}. Actinic flux, unlike solar insolation, does not refer to any specifically oriented collecting surface. This is a fundamental quantity for photochemistry, since we consider that molecules do not preferentially absorb radiation with respect to any particular orientation in space. This quantity is therefore not corrected by the cosine of the incident angle, unlike insolation. From this 3D actinic flux, the daily-averaged insolation is computed. As a comparison with Fig. \ref{fig:Actinic}, Fig. \ref{fig:Insolation} presents the daily-averaged solar insolation [W\,m$^{-2}$] received by a horizontal unit surface in Saturn's atmosphere. Following \cite{Moses2005b}, we use a solar constant of 14.97 W\,m$^{-2}$ for these calculations. In both figures, the effect of Saturn's elliptical orbit is obvious. Since Saturn reaches its perihelion shortly after the summer solstice, the amount of solar flux is more important at this time. The dark blue areas in the winter hemispheres indicate polar nights. 

Ring shadowing effects due to the A-B-C rings and to the Cassini division are also included. \citet{Brinkman1979} and \citet{Bezard1986} have first calculated ring shadowing in atmospheric models, however we adopt the prescription of \citet{Guerlet2014}, which is more suited for implementation in our photochemical model. This method calculates whether or not a point on the planet at a given latitude and longitude is under the shadow of the rings. If this is the case, the solar flux at this point is reduced by the ring opacity. We adopt the normal opacity profile of the rings from \citet{Guerlet2014}, which is based on more than 100 stellar occultations measured by the UVIS instrument aboard Cassini \citep{Colwell2010}. Finally, these normal opacities are corrected to account for the incidence angle of radiation over the rings. Diffusion effects from the ring are not included. We account for the latitudinal extent of the numerical cells in our calculations. Here we present results from simulations that use 10$^{\circ}$-wide latitudinal cells. Therefore, the ring occultation is averaged over these 10$^{\circ}$-wide cells. Each of the 10$^{\circ}$-wide cells have been sampled over 0.1$^{\circ}$-wide sub-cell.
The effect of ring shadowing can be seen at mid latitudes in the winter hemispheres in Figs. \ref{fig:Actinic} and \ref{fig:Insolation}.

%

\section{Results}
\label{section:results}

In this section, we first present results from our photochemical model using the spatially uniform thermal field (U) described in \ref{subsubsection:Uniform} in order to compare with existing models \citep{Moses2005b}. We detail the variability in hydrocarbon abundances as a function of altitude/latitude/time only due to the variation of the heliocentric distance of Saturn and of the latitude of the sub-solar point. Then, we give a brief overview of the influence of the rings on chemistry. Finally, we present the effect induced by the seasonal temperature field (S) and compare the results with the spatially uniform case. The interest here lies in the fact that we first present photochemical results using a simple test-case, i.e. a spatio-temporally uniform case previously studied \citep{Moses2005b}, before adding more complexity by considering a more realistic thermal field.

\subsection{Seasonal variability with the spatially uniform thermal field}
\label{subsection:Seasonnal_Variability}

We present here the results from seasonal simulations using the spatially uniform (U) temperature field, with an emphasis on methane, ethane and acetylene as they are the most important compounds with respect to the radiative heating/cooling of the atmosphere \citep{Yelle2001}.

\subsubsection{Methane (CH$_4$), ethane (C$_2$H$_6$) and acetylene (C$_2$H$_2$)}

\begin{figure}[htp]
\begin{center}
\includegraphics[width=1.0\columnwidth]{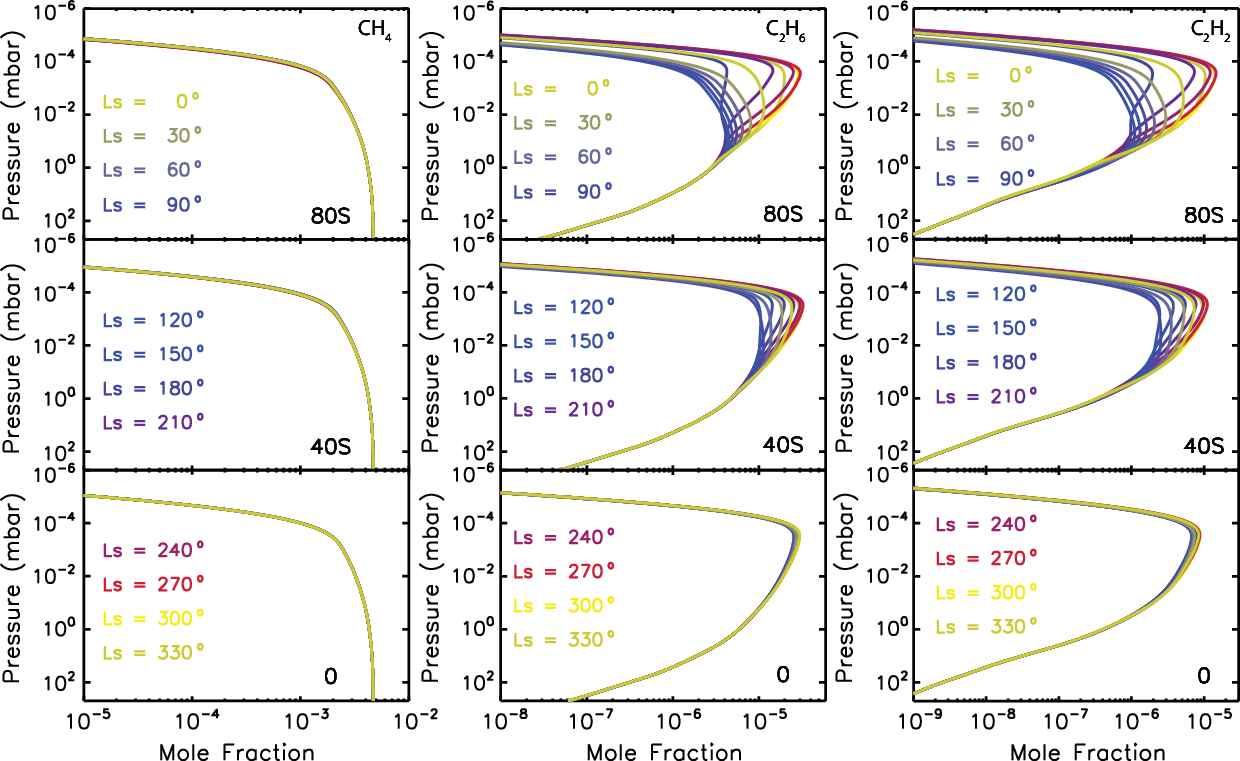}
\caption{Seasonal evolution of CH$_4$ (left), C$_2$H$_6$ (center) and C$_2$H$_2$ (right) vertical profiles computed for the whole Kronian year (360$^{\circ}$ in heliocentric longitude using a 30$^{\circ}$ step). Three latitudes are presented: 80$^{\circ}$S (top), 40$^{\circ}$S (middle) and the equator (bottom). CH$_4$ does not show any strong seasonal variability, even at high latitudes. Its shape is controlled by vertical mixing rather than photolysis \citep{Romani1988}. The seasonal variability on C$_2$H$_6$ and C$_2$H$_2$ is clearly seen at low-pressure levels and high latitudes due to the large insolation variation at such latitudes over one Kronian year. The (U) thermal field has been used for these calculations.}
\label{fig:Seasonnal_U}
\end{center}
\end{figure}

The vertical profiles of CH$_4$, C$_2$H$_6$, and C$_2$H$_2$, using the spatially uniform temperature field, are displayed in Fig. \ref{fig:Seasonnal_U}.

CH$_4$ does not exhibit strong seasonal variations, as eddy mixing and molecular diffusion, rather than photolysis, are the major processes controlling the shape of its vertical distribution \citep{Romani1988}. Indeed, due to its relatively high abundance, CH$_4$ is never depleted enough to show seasonal variations.

The seasonal variability on C$_2$H$_6$ is clearly seen at low-pressure levels. The shape of its vertical profile is mostly governed by reaction R20 (CH$_3$ + CH$_3$  $\rightarrow$ C$_2$H$_6$). Methyl is produced from the CH$_4$ photolysis around 10$^{-4}$\,mbar. At higher pressures, C$_2$H$_6$ is mostly in diffusive equilibrium (see \citealt{Zhang2013} for instance) and its shape is governed by the slow diffusion of C$_2$H$_6$ produced at lower pressure levels. The seasonal variability of this compound is correlated with insolation and is therefore maximum around the poles.

C$_{2}$H$_{2}$ shows a seasonal variability similar to C$_2$H$_6$, with some differences arround the 1\,mbar pressure level, because C$_2$H$_2$ has substantial production at this pressure level by reactions R6 (H + C$_2$H$_3$  $\rightarrow$ C$_2$H$_2$ + H$_2$) and R21 (C$_2$H + H$_2$ $\rightarrow$ C$_2$H$_2$ + H), and depletion by R5 (H + C$_2$H$_2$  $\rightarrow$ C$_2$H$_3$) and R48 (C$_2$H$_2$ + $h \nu$  $\rightarrow$ C$_2$H + H).

\subsubsection{Evolution of the C$_2$H$_2$ column density}

\begin{figure}[htp]
\begin{center}
\includegraphics[width=0.45\columnwidth]{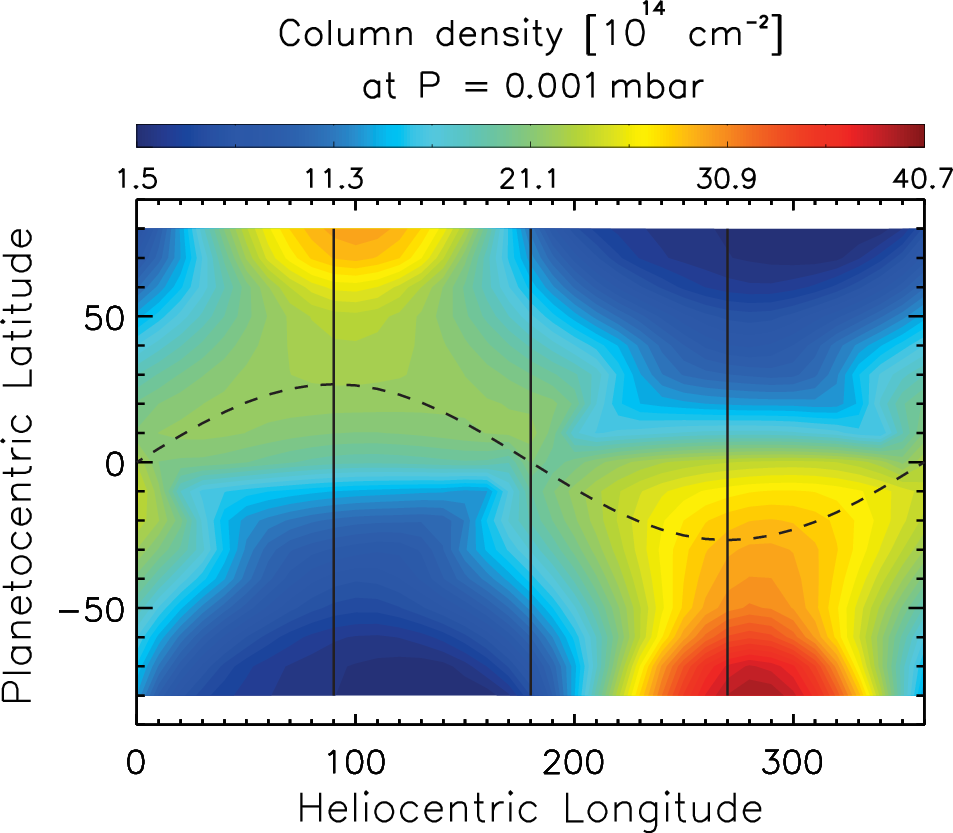}
\includegraphics[width=0.445\columnwidth]{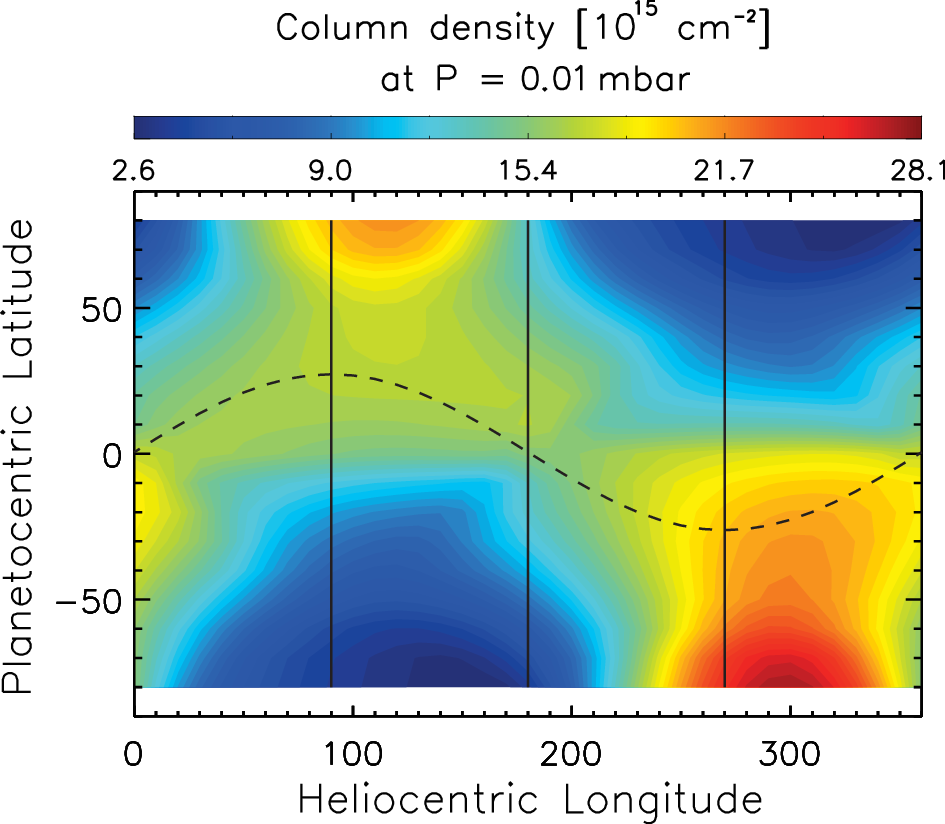}\\
\includegraphics[width=0.45\columnwidth]{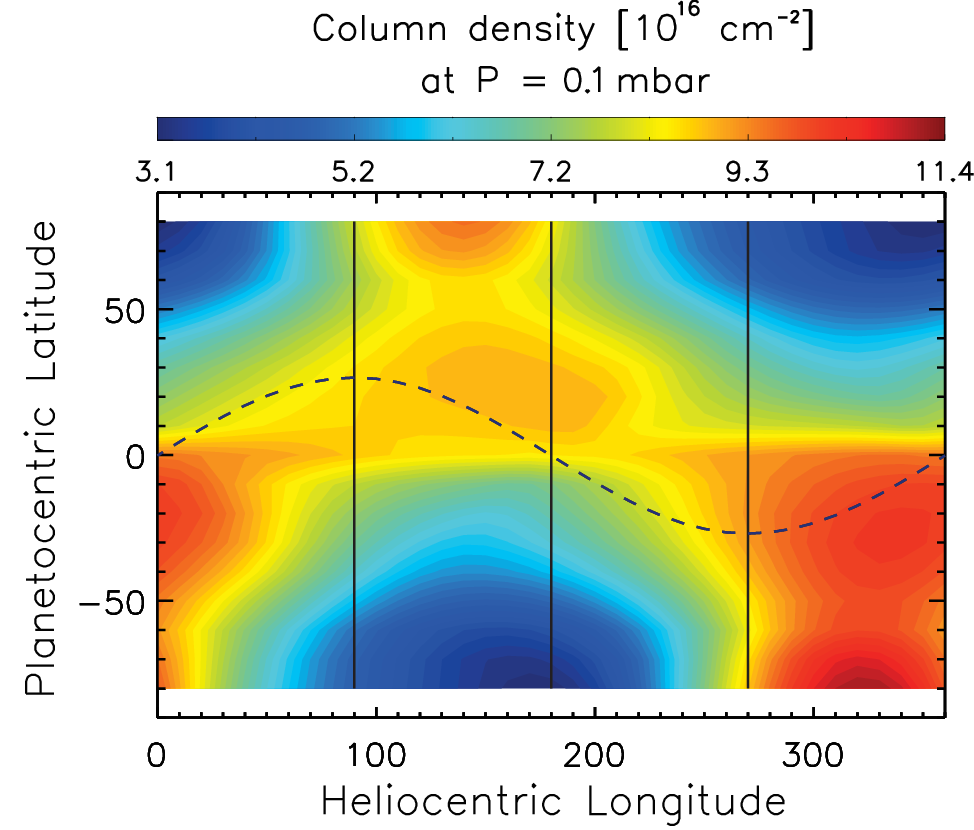}
\includegraphics[width=0.432\columnwidth]{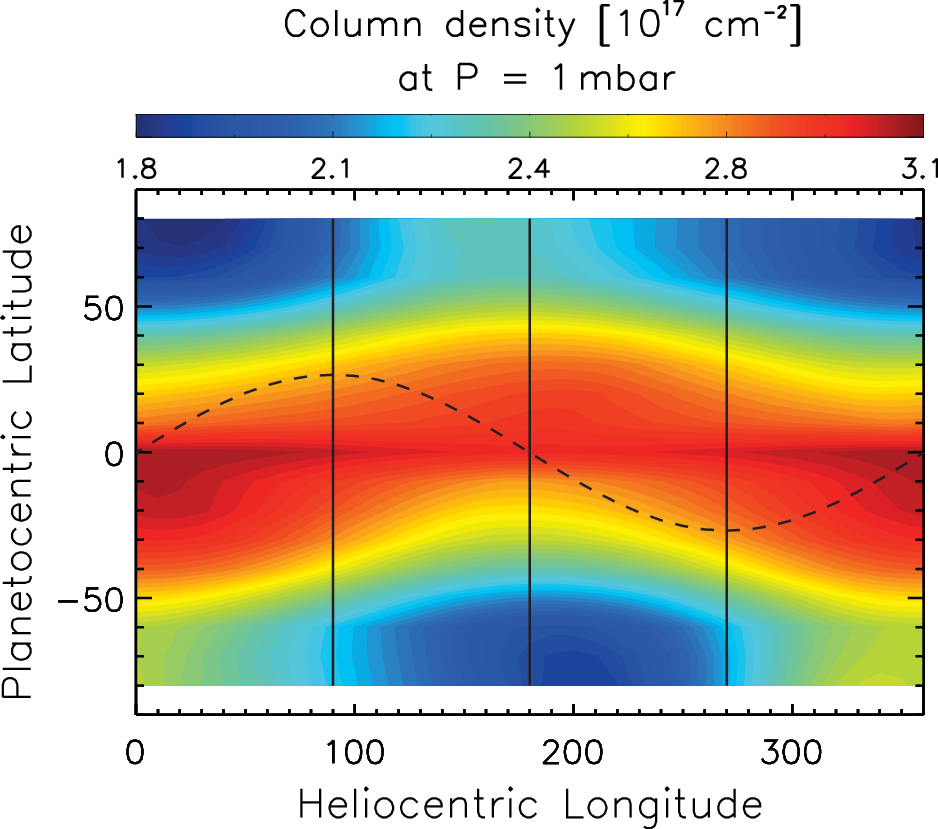}
\caption{C$_2$H$_2$ column density [cm$^{-2}$] above 10$^{-3}$\,mbar (top left), 10$^{-2}$\,mbar (top right), 0.1\,mbar (bottom left), and 1\,mbar (bottom right). Ring shadowing is clearly seen around the winter solstice in the winter hemisphere. The solid line indicate the solstices and equinoxes, while the dashed line indicates the position of the subsolar point along the year. The North-South asymmetry between the summer hemispheres is caused by Saturn's eccentricity, its perihelion occurring shortly after the southern summer solstice. In both summer hemispheres, after the summer solstice, the C$_2$H$_2$ column densities reach a maximum which is shifted in time with respect to the maximum insolation level, i.e. at the solstice itself.}
\label{fig:Seasonnal_coldens_C2H2}
\end{center}
\end{figure}

The C$_2$H$_2$ column densities computed for pressures lower than 10$^{-3}$\,mbar, 10$^{-2}$\,mbar, 0.1\,mbar and 1\,mbar are presented in Fig. \ref{fig:Seasonnal_coldens_C2H2} , as a comparison with Fig. 8 of \citet{Moses2005b}. The results concerning the temporal evolution of this column density are in good agreement. The differences in the column density absolute values can be attributed to differences in the temperature/pressure background, the eddy diffusion profile or the chemical network.
At 10$^{-3}$\,mbar, an asymmetry between northern and southern summer solstices is caused by Saturn's eccentricity. The maximum value in column density is reached around the southern summer pole, shortly after the southern summer solstice, at $L_s \approx$ 280$^{\circ}$. The signature of the rings is clearly visible at low latitudes near the solstices in the winter hemisphere. The amount of radicals (and therefore chemical compound produced from radicals) is reduced (see for instance \citealt{Edgington2012}) due to the partial absorption of the UV radiation by Saturn's rings. At higher pressure levels, the ring signature is damped, and disappears almost completely at 0.1\,mbar. From that pressure level to higher ones, the abundance of C$_2$H$_2$ is mainly controlled by the downward diffusion of C$_2$H$_2$ produced at lower pressure levels. Therefore, from that pressure level, the column density features (e.g., maxima and minima) are increasingly phase-lagged with increasing pressure. These plots also show that the maximum value of the C$_2$H$_2$ column density is shifted from high latitudes to equatorial latitudes with increasing pressure in agreement with previous work of \cite{Moses2005b}. Indeed, this column density mimics the seasonal solar actinic flux at high altitudes (around 10$^{-3}$\,mbar), while it follows the annually averaged actinic flux at lower altitudes (at 1\,mbar and below) where the column density is maximum at the equator.


\subsubsection{Other species}

\begin{figure}[htp]
\begin{center}
\includegraphics[width=0.4\columnwidth]{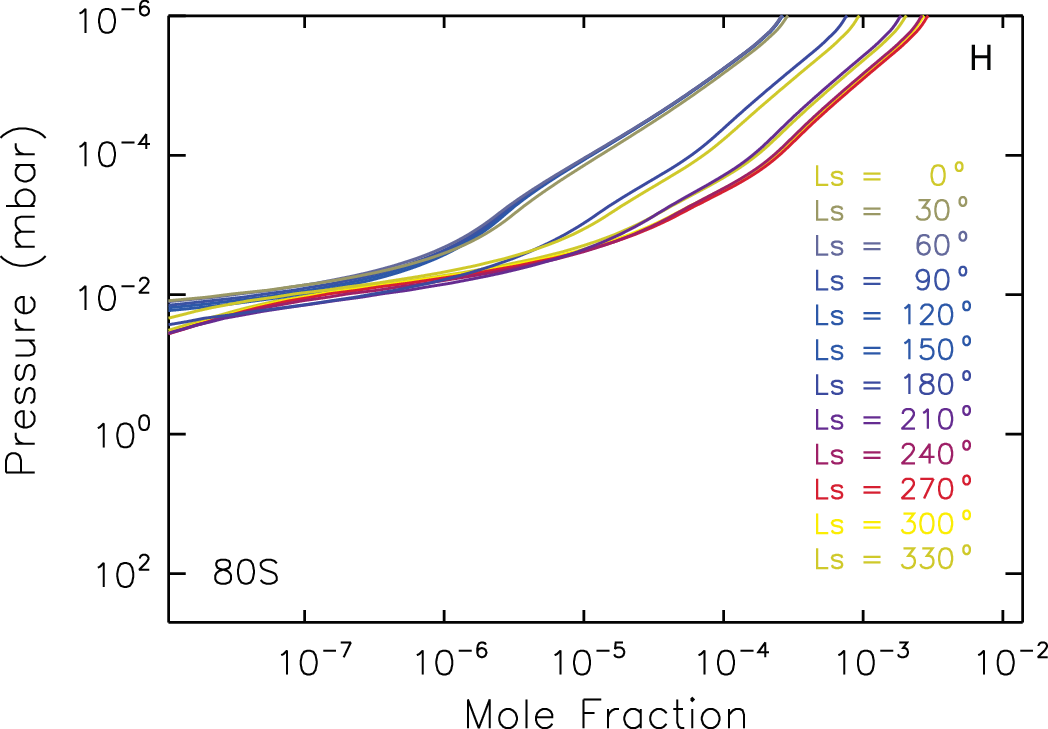}
\includegraphics[width=0.4\columnwidth]{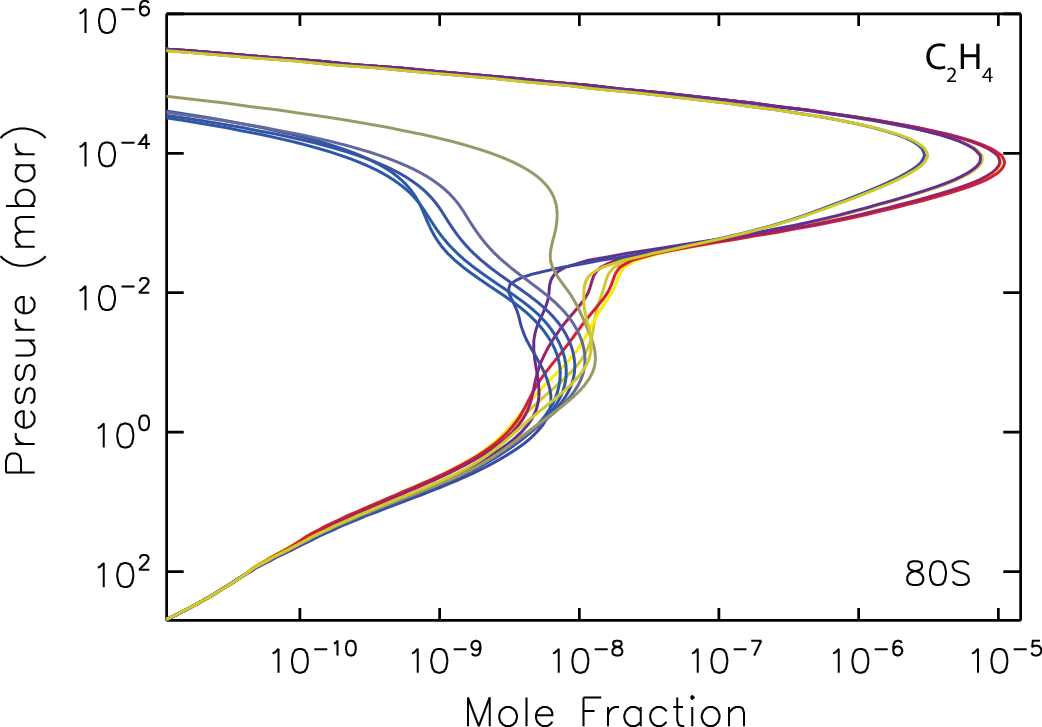}
\includegraphics[width=0.4\columnwidth]{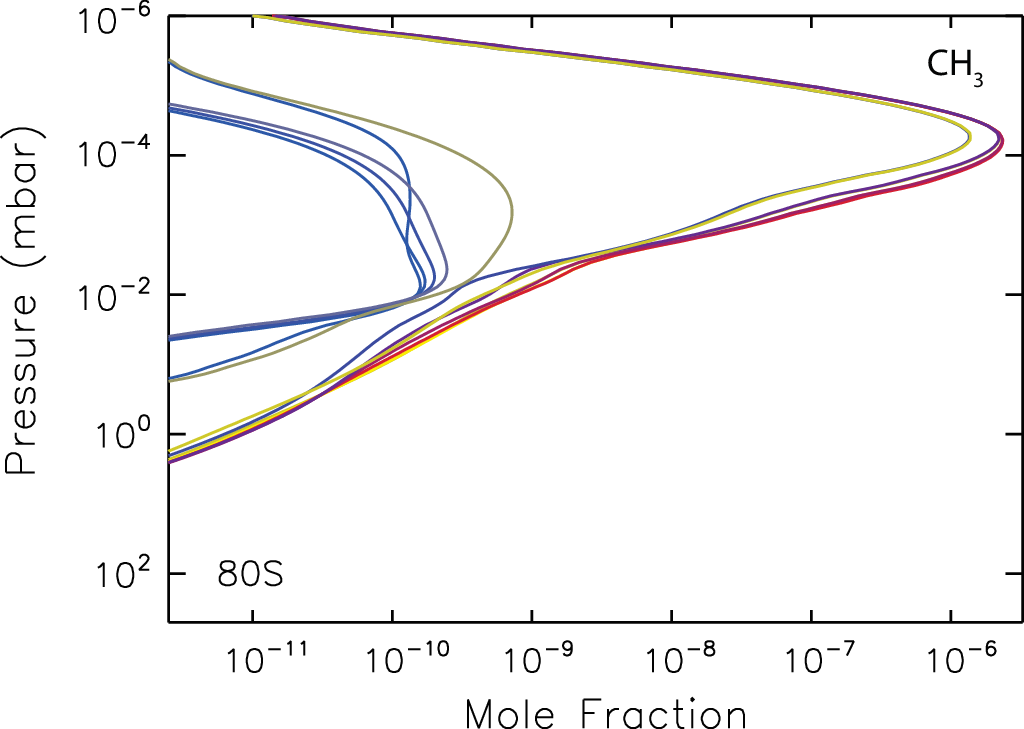}
\caption{Seasonal evolution of the mole fraction of atomic hydrogen (H, top left), ethylene (C$_2$H$_4$, top right), methyl (CH$_3$, bottom) as a function of pressure and heliocentric distance ($L_s$) at 80$^{\circ}$S. The profiles are presented using a 30$^{\circ}$ step in $L_s$.}
\label{fig:Seasonnal_H_CH3_C2H4}
\end{center}
\end{figure}

The seasonal evolution of the vertical profiles of several other compounds of interest are presented in Fig. \ref{fig:Seasonnal_H_CH3_C2H4} for 80$^{\circ}$S, where variability is expected to be most noticeable. Radicals, such as atomic hydrogen (H) and methyl (CH$_3$), show a strong seasonal variability, as they mainly result from the photolysis of CH$_4$ and depend therefore on insolation conditions. These short-lived radicals undergo a drastic decrease in their abundances in winter conditions at this latitude, i.e., when CH$_4$ photolysis is stopped by the polar night. Ethylene (C$_{2}$H$_{4}$) also shows significant seasonal changes around 10$^{-4}$\,mbar as they mainly result from reactions involving CH radicals and methane (R13: CH + CH$_4$  $\rightarrow$ C$_2$H$_4$ + H). Below that level, C$_{2}$H$_{4}$ production rate through reaction R7 (H + C$_2$H$_3$  $\rightarrow$ C$_2$H$_4$) becomes increasingly important, consistently with \citet{Moses2005b}, to be its main production process around 1\,mbar.

\subsubsection{Impact of the rings on chemistry}
\label{subsubsection:Ring_Impact}

\begin{figure}[htp]
\begin{center}
\includegraphics[width=0.8\columnwidth]{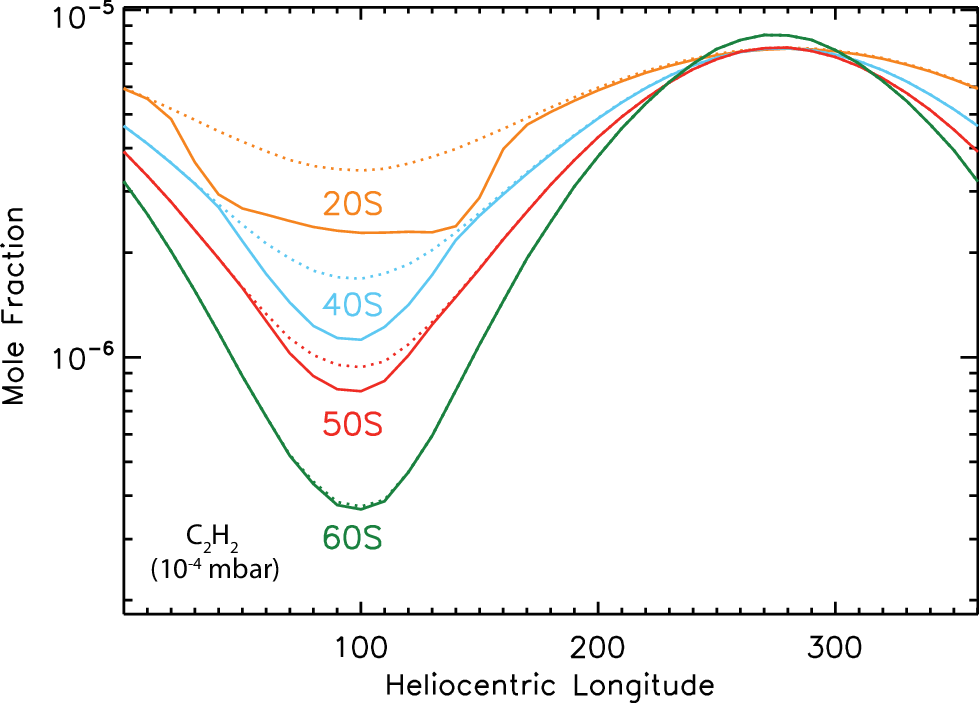} \\
\includegraphics[width=0.8\columnwidth]{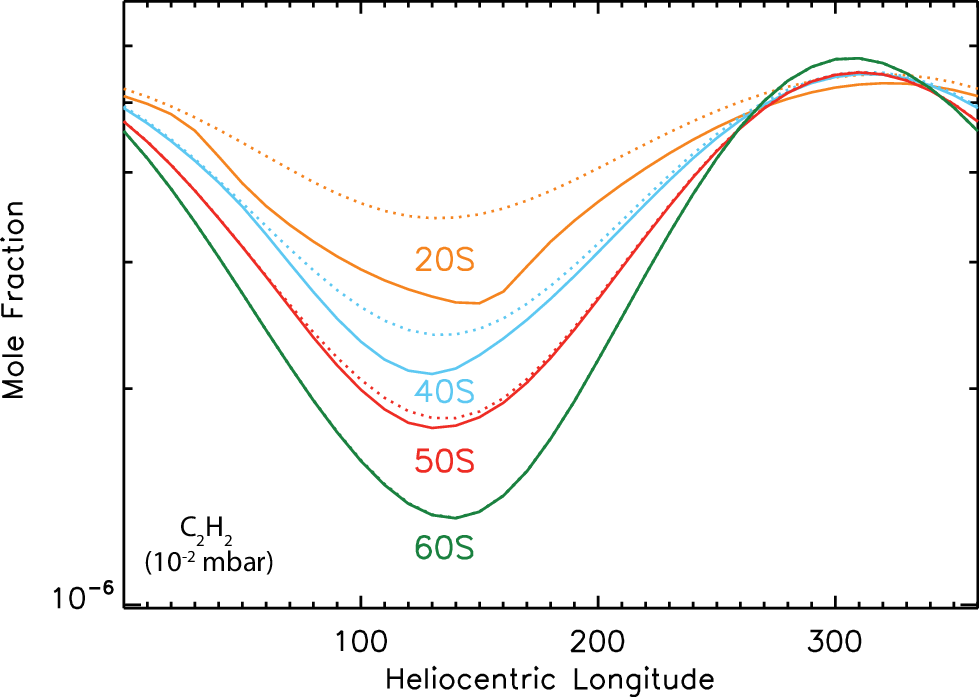}
\caption{C$_2$H$_2$ mole fraction at 10$^{-4}$\,mbar (top) and 10$^{-2}$\,mbar (bottom) as a function of heliocentric longitude. The solid lines include ring-shadowing effects, whereas the dotted lines do not include this effect. These effects are only visible from the equator to $\pm$50$^{\circ}$. High latitudes are alternately in polar day and polar night. The mole fraction minima and maxima are damped and phase-lagged at 10$^{-2}$\,mbar with respect to 10$^{-4}$\,mbar.}
\label{fig:Rings1}
\end{center}
\end{figure}

\begin{figure}[htp]
\begin{center}
\includegraphics[width=0.8\columnwidth]{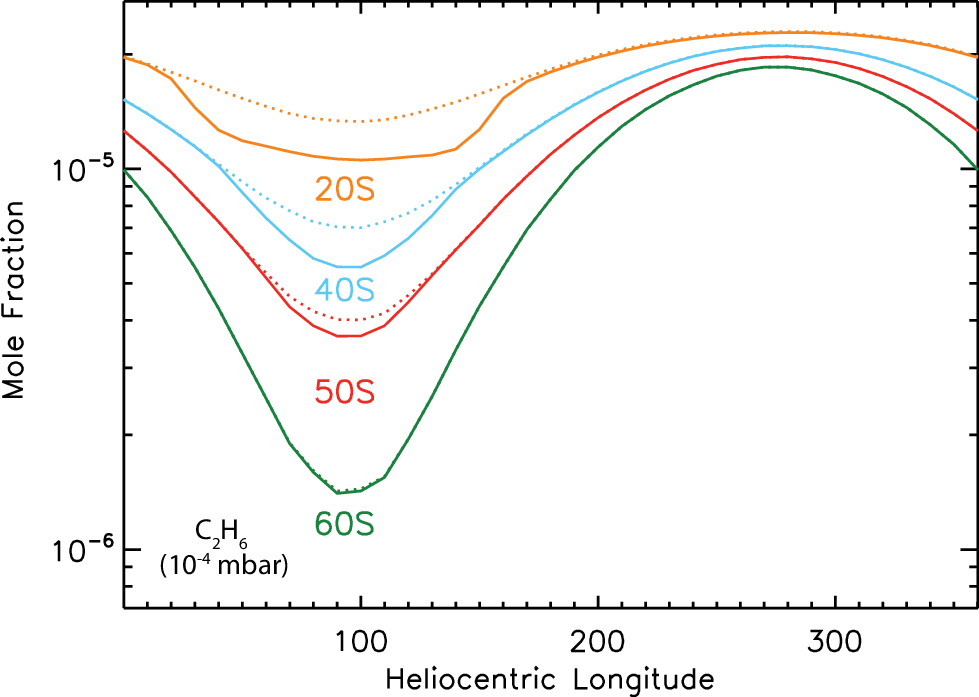}
\includegraphics[width=0.8\columnwidth]{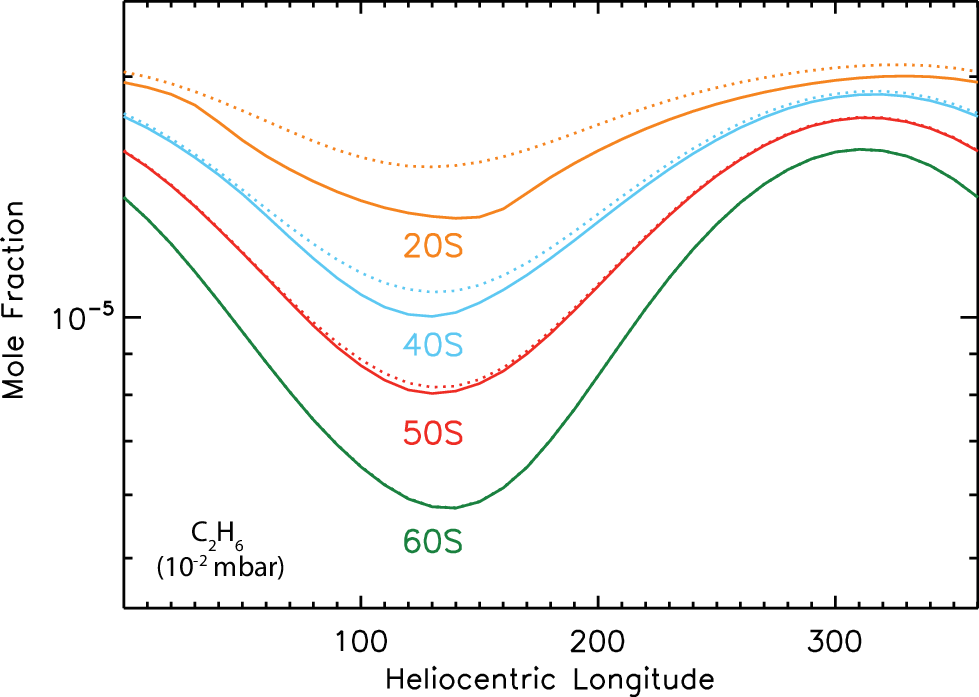}
\caption{Same as Fig. \ref{fig:Rings1} for C$_2$H$_6$. Solid and dotted lines represent photochemical predictions with and without ring shadowing, respectively.}
\label{fig:Rings2}
\end{center}
\end{figure}

The impact of the UV absorption by the rings on the seasonal evolution of C$_2$H$_2$ and C$_2$H$_6$ mole fractions are depicted in Fig. \ref{fig:Rings1} and \ref{fig:Rings2}. 
As expected from geometrical considerations, and given the latitudinal extent of the numerical cells of the model, the ring shadowing effect is maximum at latitudes below 50$^{\circ}$ in the winter hemisphere, at the solstice itself, i.e. when the shadow cast by the ring on the planet are the most extended in that hemisphere. The impact of the ring shadowing is more localised in time at a latitude of 40$^{\circ}$ than at a latitude of 20$^{\circ}$ in the winter hemisphere. At these latitudes, the main impact on chemistry of the ring shadowing effect comes from Saturn B ring. At a latitude of 20$^{\circ}$ in the winter hemisphere, the ring shadowing effects are effective over 140$^{\circ}$ in L$_s$, while at a latitude of 40$^{\circ}$, they are effective over 80$^{\circ}$ in L$_s$. At higher pressure levels, and similarly to the column density, the mole fraction minima and maxima are damped and phase-lagged.

\subsection{Accounting for the seasonal temperature field}
\label{subsection:Seasonnal_Variability2}

\begin{figure}[htp]
\begin{center}
\includegraphics[width=1.0\columnwidth]{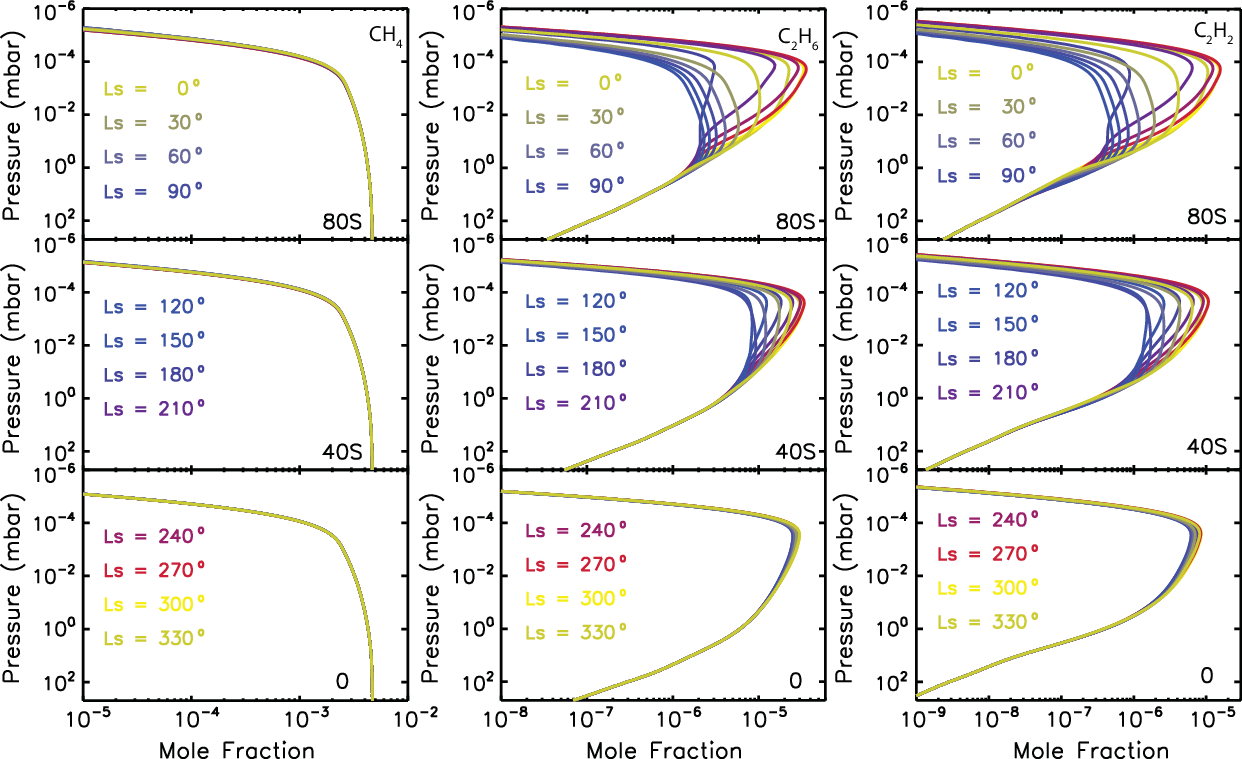}
\caption{Seasonal evolution of CH$_4$ (left), C$_2$H$_6$ (center) and C$_2$H$_2$ (right) vertical profiles computed for the whole Kronian year (360$^{\circ}$ in heliocentric longitude using a 30$^{\circ}$ step) and using the seasonal (S) thermal field.}
\label{fig:Seasonnal_S}
\end{center}
\end{figure}

In this section, we present results from the seasonal simulations using the seasonal (S) thermal profiles. The vertical profiles of CH$_4$, C$_2$H$_6$, and C$_2$H$_2$, using this thermal field are displayed in Fig. \ref{fig:Seasonnal_S}. These profiles have to be compared with Fig. \ref{fig:Seasonnal_U}, where the (U) thermal field was used.

Taking the (S) field into account leads to differences with respect to the (U) case in the amplitude of the seasonal variability of C$_2$H$_2$ and C$_2$H$_6$ at pressure levels ranging from 10$^{-5}$ to 10$^{-1}$\,mbar. C$_2$H$_2$ now shows a small seasonal variability at pressure levels ranging from 0.5 to 10\,mbar, which was not the case previously. The position of the homopause is also expected to vary, as the molecular diffusion coefficient has a thermal dependency ($D_i$ $\propto$ $T^{1.75}/p$). Using the (S) field, the homopause is generally shifted to a lower pressure, due to the fact that the (U) thermal field corresponds to summer conditions at latitude of 20$^{\circ}$ in the summer hemisphere.

\begin{figure}[htp]
\begin{center}
\includegraphics[width=0.6\columnwidth]{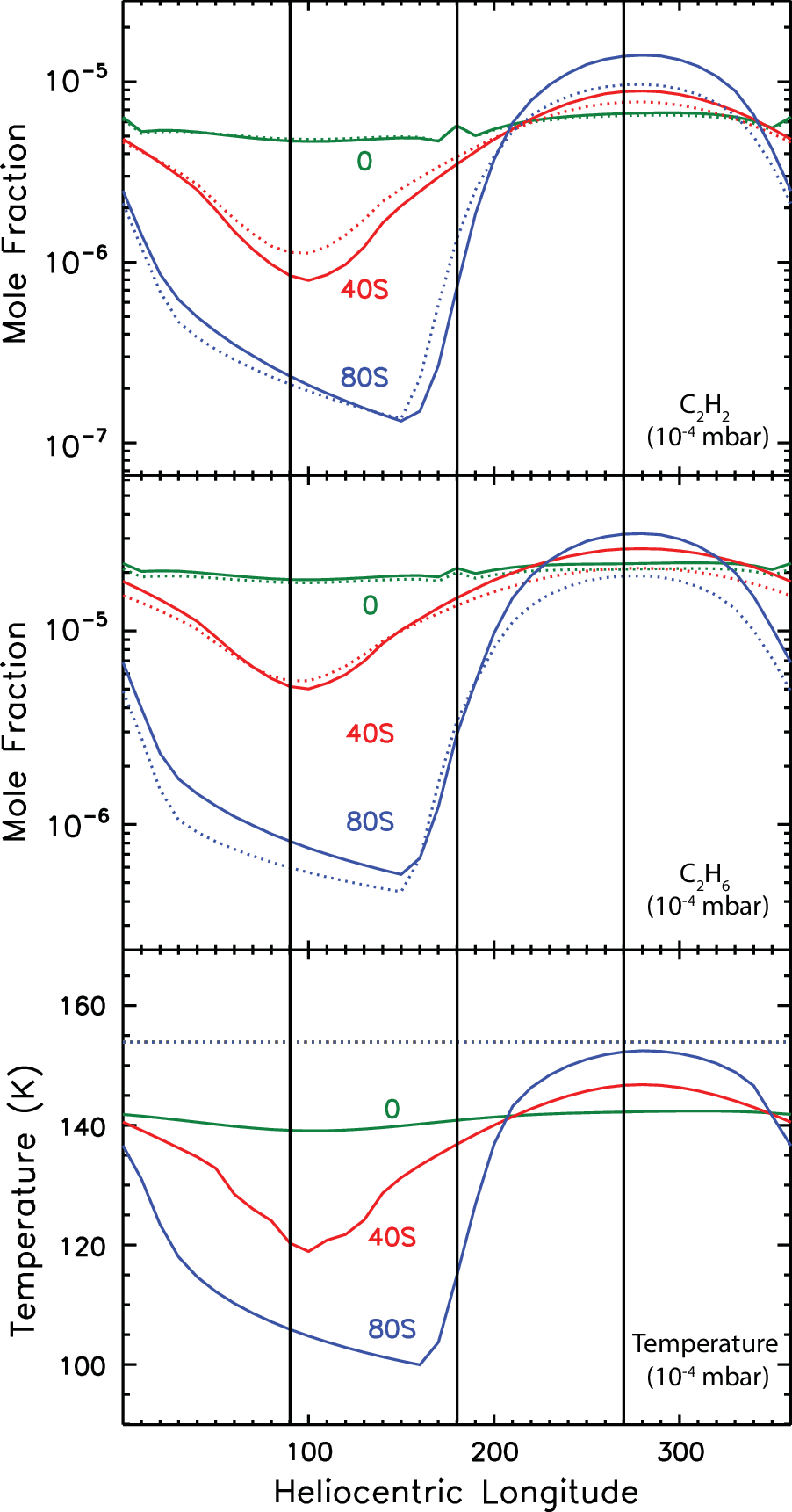}
\caption{Evolution of the C$_2$H$_2$ mole fraction (top panel), the C$_2$H$_6$ mole fraction (middle panel) and temperature (bottom panel) at 10$^{-4}$\,mbar. The evolution of mole fractions and temperature is plotted as a function of heliocentric longitude for latitudes of 80$^{\circ}$S, 40$^{\circ}$S and at the equator. Values obtained with the (S) and (U) thermal fields are displayed with solid and dotted lines, respectively. The black solid lines indicate the position of the solstices and equinoxes (see Fig. \ref{fig:Seasons}).}
\label{fig:Seasonnal_Comparaison_1}
\end{center}
\end{figure}

The seasonal evolutions of the C$_2$H$_2$ and C$_2$H$_6$ mole fractions at three pressure levels (10$^{-4}$, 10$^{-2}$ and 1\,mbar) and considering the (S) and (U) thermal fields are shown in Figs. \ref{fig:Seasonnal_Comparaison_1}, \ref{fig:Seasonnal_Comparaison_2} and \ref{fig:Seasonnal_Comparaison_3}. For the sake of comprehension, the seasonal evolution of temperatures at these same pressure levels are shown alongside as well as the position of the different solstices and equinoxes. We only present these seasonal evolutions at a few latitudes in the southern hemisphere, although the same occurs in the northern hemisphere. Therefore, in what follows, summer and winter refer to these seasons in the southern hemisphere, if not otherwise specified.

$\bullet$ \textbf{Fig. \ref{fig:Seasonnal_Comparaison_1}:} At 10$^{-4}$\,mbar, C$_2$H$_2$ and C$_2$H$_6$ mole fractions, as predicted using both (S) and (U) thermal fields, evolve in phase. Around the summer solstice (L$_s$ = 270$^{\circ}$), the abundance of these compounds is increased when considering the (S) thermal field and they both show a positive abundance gradient from the equator to the south pole. Note that, when considering the (U) field, C$_2$H$_6$ shows a very small abundance gradient at the summer solstice.

The differences in C$_2$H$_2$ and C$_2$H$_6$ abundances between both (S) and (U) thermal field calculations never exceed 50\% except at high latitudes at summer solstice where C$_2$H$_2$ and C$_2$H$_6$ abundances are enhanced by a factor of 1.3 and 1.6, respectively. The small bump observed at the equator with both thermal fields is due to the absence of ring shadowing due to the thin nature of the rings. The ring opacity on the UV field is averaged over the latitudinal extent of the numerical cells, which are 10$^{\circ}$-wide here. A local maximum on the UV field is expected at the equator at L$_s$ = 0$^{\circ}$ and 180$^{\circ}$, i.e. when the projection of the rings over Saturn's planetary disk is negligible, given the latitudinal extent of the numerical cells. The temperatures at 40$^{\circ}$S for L$_s$ ranging from 50$^{\circ}$ to 140$^{\circ}$ vary abruptly with times due to the shadowing from the different rings. The temperature of the (U) thermal field at this pressure level is 1\,K warmer than the one of (S) thermal field at latitude of 80$^{\circ}$S around the summer solstice.

\begin{figure}[htp]
\begin{center}
\includegraphics[width=0.45\columnwidth]{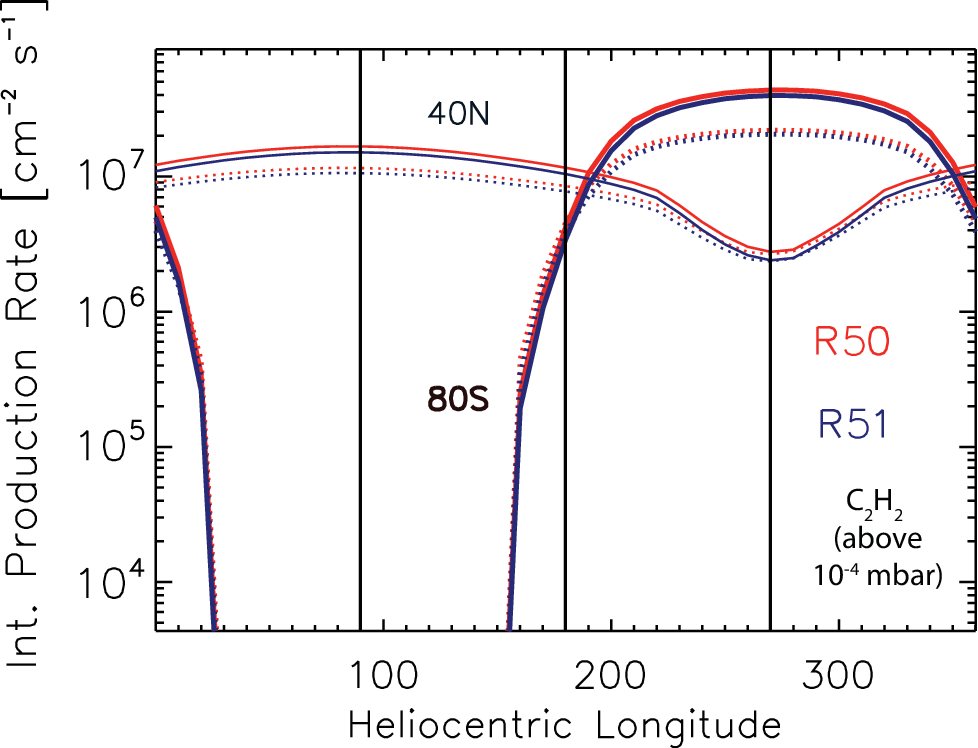}
\includegraphics[width=0.46\columnwidth]{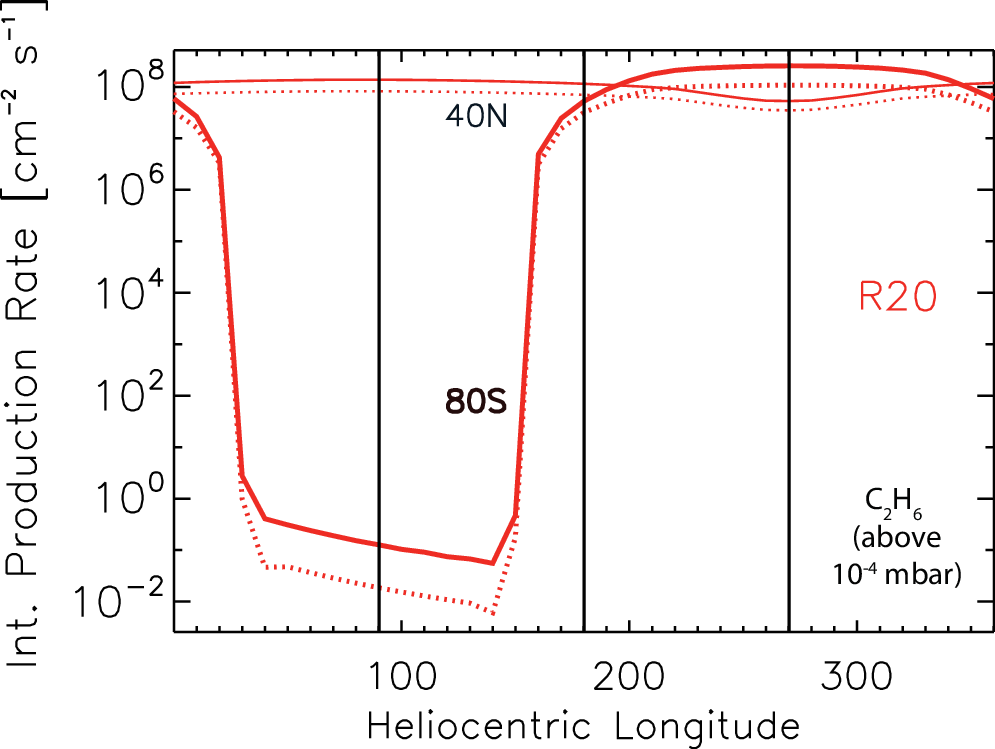}
\caption{Seasonal evolution of the integrated production rate above 10$^{-4}$ mbar of the main reactions leading to the production of C$_2$H$_2$ (left panel) and C$_2$H$_6$ (right panel). For the sake of clarity these integrated production rates are presented at 80$^{\circ}$S (thick lines) and 40$^{\circ}$N (thin lines). Calculations that include the (S) and (U) thermal field are displayed with solid and dotted lines, respectively.}
\label{fig:Integ_Prod_1}
\end{center}
\end{figure}

\begin{figure}[htp]
\begin{center}
\includegraphics[width=0.6\columnwidth]{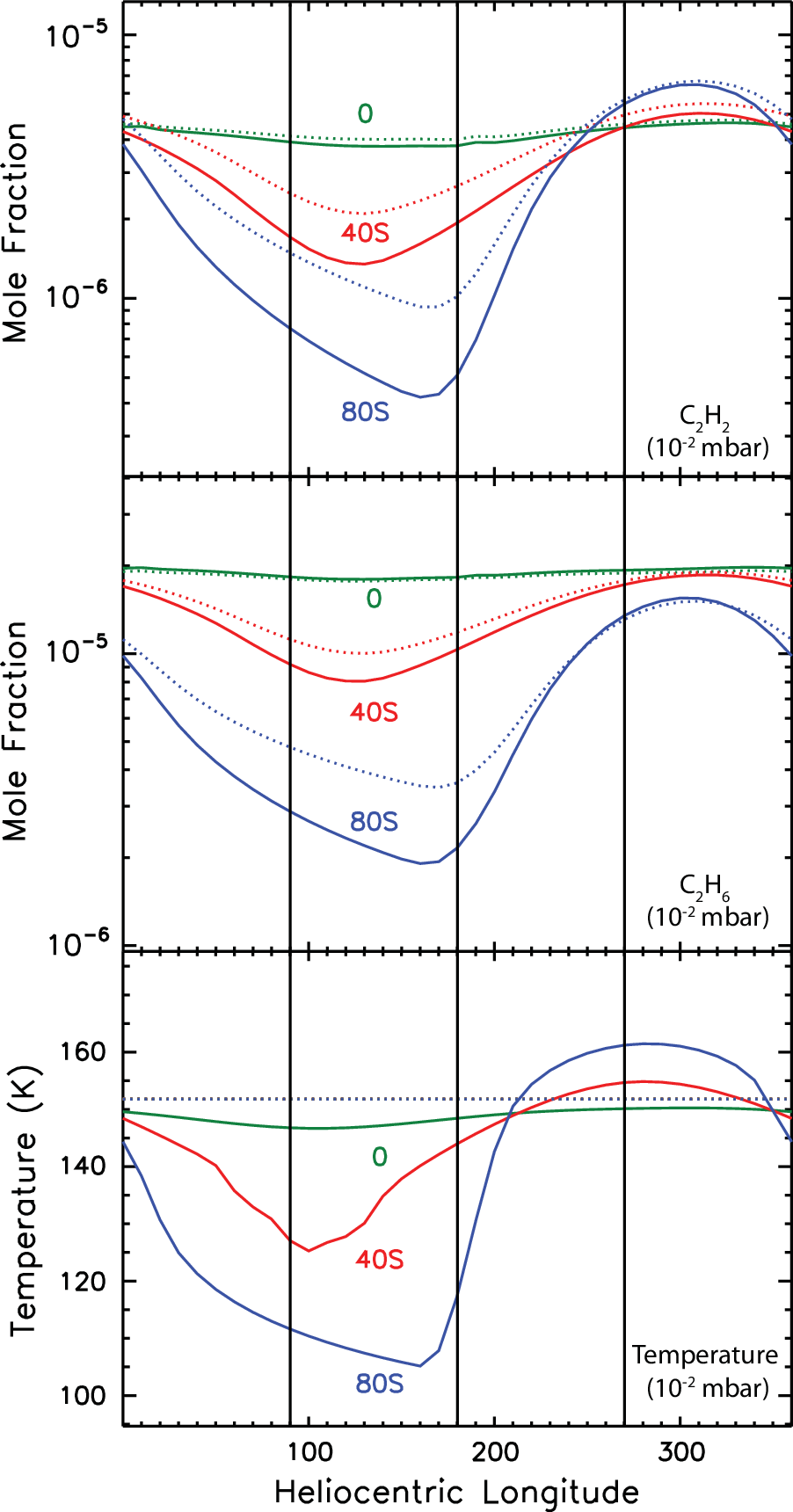}
\caption{Same as Fig. \ref{fig:Seasonnal_Comparaison_1} at 10$^{-2}$\,mbar.}
\label{fig:Seasonnal_Comparaison_2}
\end{center}
\end{figure}

At 10$^{-4}$\,mbar, the C$_2$H$_2$ production is mainly controlled by reactions R50 (C$_2$H$_4$ + $h \nu$ $\rightarrow$ C$_2$H$_2$ + H$_2$) and R51 C$_2$H$_4$ + $h \nu$ $\rightarrow$ C$_2$H$_2$ + H + H) as displayed on Fig. \ref{fig:Integ_Prod_1} (left panel). The integrated production rates above that pressure level computed using the (S) thermal field are always greater than the ones computed with the (U) field. These differences are caused by the temperature which affects the position of the homopause and allows the UV radiation to penetrate deeper in the (U) thermal field case. This ultimately leads to an increase in the integrated production rate above 10$^{-4}$\,mbar of CH radical from methane photolysis. Since the (U) thermal field is hotter than the (S) thermal field at all times along the year, the integrated production rates of C$_2$H$_2$ and C$_2$H$_6$ above 10$^{-4}$\,mbar in the former case are always expected to be greater than the (U) case. From this radical, C$_2$H$_4$ is produced through reaction R13, and then photolysed through reactions R50 and R51. We note that the differences in the integrated production rates between these two thermal fields reach a minimum at 40$^{\circ}$N around the northern winter solstice (L$_s$ = 270$^{\circ}$) while, at the same time, the C$_2$H$_2$ mole fraction becomes more important when considering the (U) thermal field than when using the (S) thermal field. These differences are produced by the decrease in the diffusion timescale due to the contraction of the atmospheric column which cools down around the winter solstice as explained below.

A similar behavior is observed for C$_2$H$_6$ (Fig. \ref{fig:Integ_Prod_1}, right panel) whose integrated production rates are controlled by reaction R20. We can however note that, at 80$^{\circ}$S and around the winter solstice (L$_s$ = 90$^{\circ}$) the integrated production rate considering the (S) thermal field is more important than the one using (U) thermal field, consistent with the predicted greater abundance of C$_2$H$_6$ at that time.


$\bullet$ \textbf{Fig. \ref{fig:Seasonnal_Comparaison_2}:} At 10$^{-2}$\,mbar, C$_2$H$_2$ is less abundant at every latitude when the (S) field is accounted for. Its abundance gradually increases with latitude from the equator to the south pole during the summer season. However, at this pressure level, the peak in the C$_2$H$_2$ and C$_2$H$_6$ abundances during summer is occuring earlier at high latitudes than at mid-latitudes due to Saturn's obliquity. C$_2$H$_6$ becomes as abundant with the (S) thermal field as it was with the (U) thermal field during the summer season. A slight dephasing is noted between the two thermal field calculations when C$_2$H$_2$ and C$_2$H$_6$ reach both their maximal and minimal values. At 80$^{\circ}$S, the abundance of C$_2$H$_2$ and C$_2$H$_6$ is decreased by a factor 2.2 and 1.7, respectively, during the winter season when we consider the (S) thermal field. At this pressure level, the temperatures of the (U) field are 2\,K and 6\,K warmer than the temperatures of the (S) field at the summer solstice and the winter solstice respectively.

\begin{figure}[htp]
\begin{center}
\includegraphics[width=0.45\columnwidth]{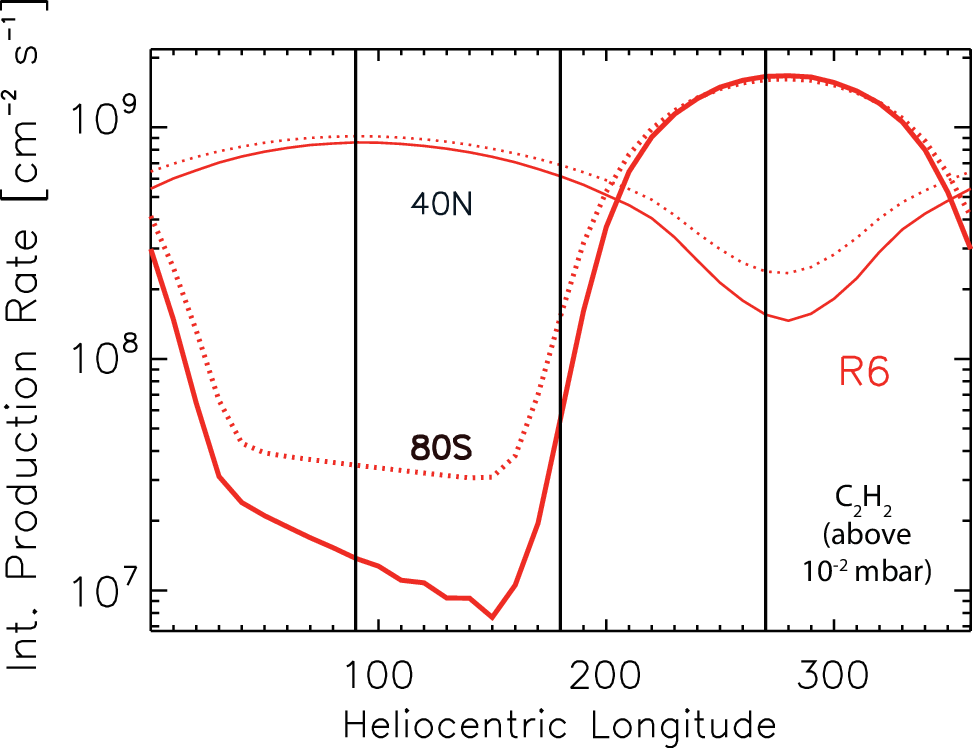}
\includegraphics[width=0.45\columnwidth]{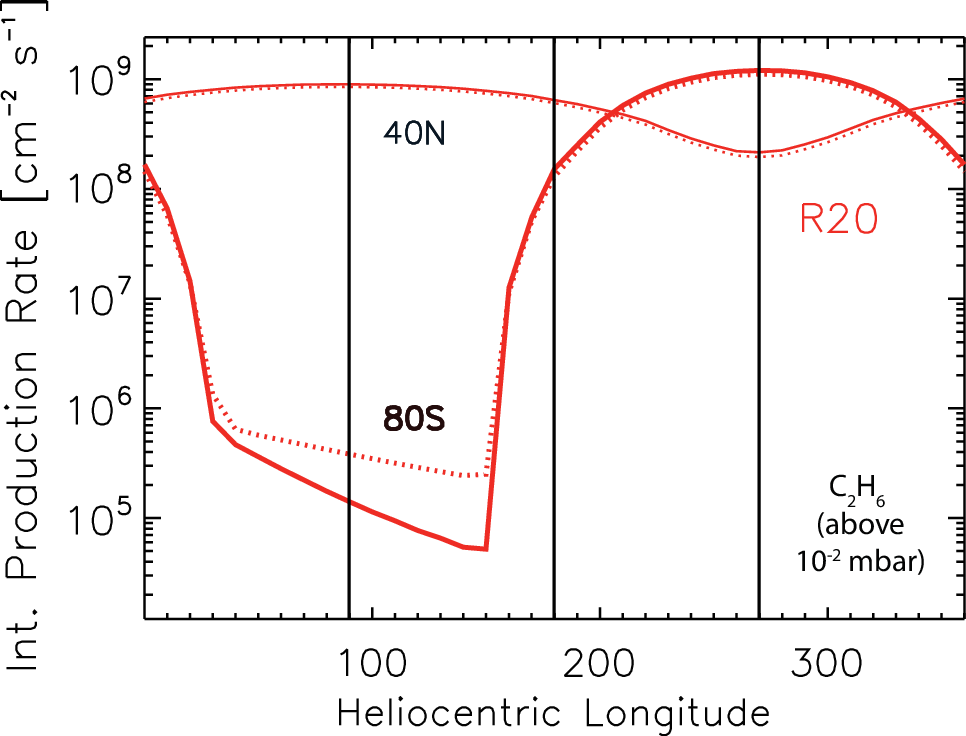}
\caption{Same as Fig. \ref{fig:Integ_Prod_1} for 10$^{-2}$ mbar.}
\label{fig:Integ_Prod_2}
\end{center}
\end{figure}

\begin{figure}[htp]
\begin{center}
\includegraphics[width=0.6\columnwidth]{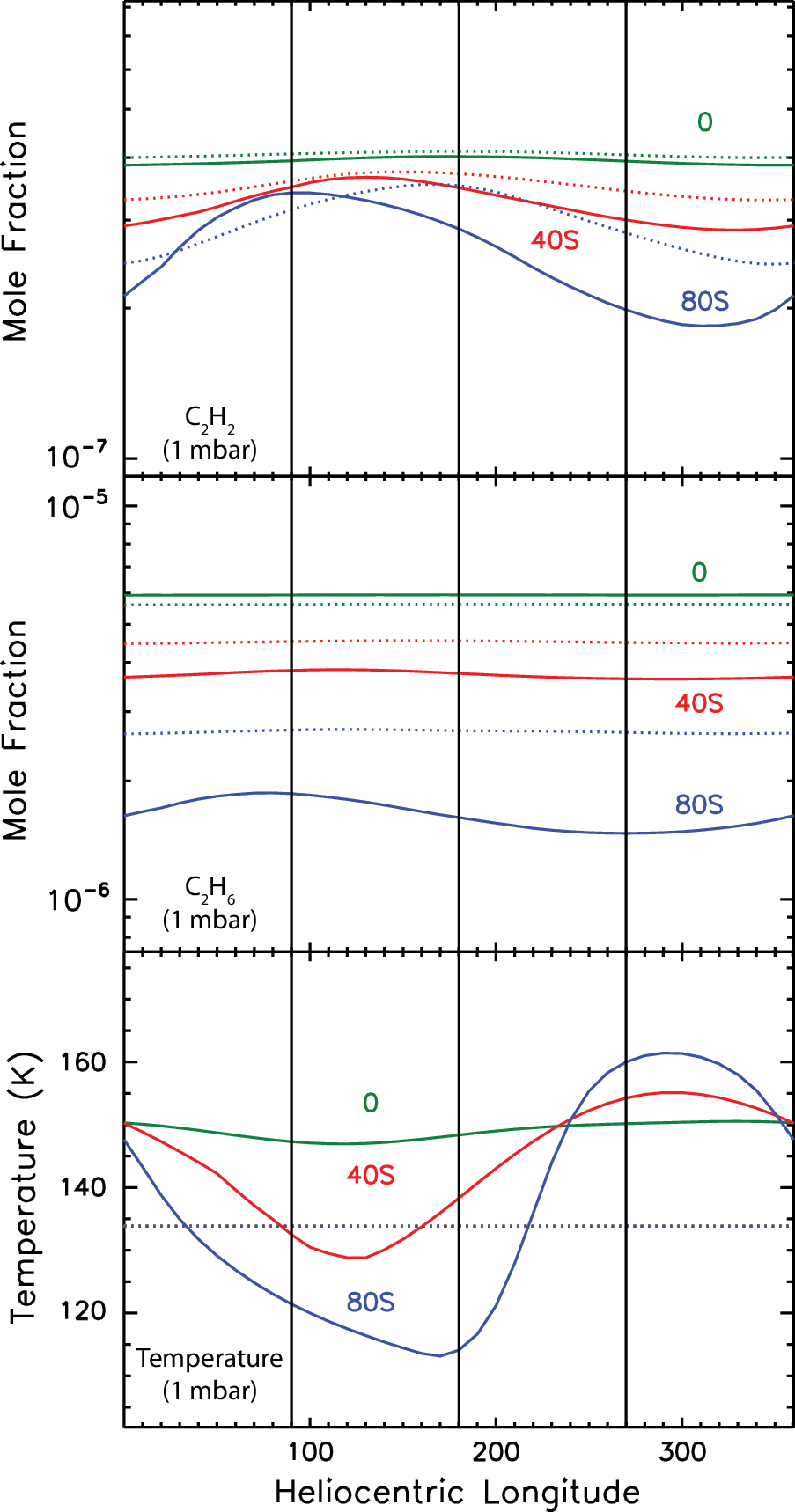}
\caption{Same as Fig. \ref{fig:Seasonnal_Comparaison_1} at 1\,mbar.}
\label{fig:Seasonnal_Comparaison_3}
\end{center}
\end{figure}

The C$_2$H$_2$ production is mainly controlled by reaction R6. The seasonal evolution of the integrated production rate of this reaction above this pressure level is presented in Fig. \ref{fig:Integ_Prod_2} (left panel) at latitudes of 80$^{\circ}$S and 40$^{\circ}$N. Around the summer solstice (L$_s$ = 270$^{\circ}$ for 80$^{\circ}$S and L$_s$ = 90$^{\circ}$ for 40$^{\circ}$N), the integrated production rate of reaction R6 in the (U) and (S) cases are very similar, consistent with the predicted C$_2$H$_2$ mole fraction. Around the winter solstice (L$_s$ = 90$^{\circ}$ for 80$^{\circ}$S and L$_s$ = 270$^{\circ}$ for 40$^{\circ}$N), the (S) integrated production rate is higher than the (U) one, also consistent with the predicted C$_2$H$_2$ mole fractions.

Similarly to the situation at lower pressure levels, the integrated C$_2$H$_6$ production rate (Fig. \ref{fig:Integ_Prod_2}, right panel) is controlled by reaction R20. The integrated production rates of this reaction are very similar between the two thermal field cases, all along the year except around the southern winter solstice at high latitudes. Indeed, the integrated production rate become less important in the (S) case than in the (U) case.

At this pressure level, the differences between the two thermal field cases observed in the C$_2$H$_2$ and C$_2$H$_6$ abundances are mainly controlled by two quantities. First, a higher integrated production rate above that level will produce a greater quantity of the considered molecule. At the same time, the contraction of the atmospheric column during the winter season will increase the diffusion of the produced hydrocarbons to higher pressure levels. This former effect is clearly noticed for C$_2$H$_6$ at 40$^{\circ}$N during the northern winter season where its integrated production rate above that level with the (S) thermal field is slightly greater than the one with the (U) thermal field, while its predicted abundance is lower in the (S) case than in the (U) case.

$\bullet$ \textbf{Fig. \ref{fig:Seasonnal_Comparaison_3}:} At 1\,mbar, C$_2$H$_2$ still shows seasonal variability while C$_2$H$_6$ seasonal variability is negligible with the (U) thermal field. The variability of these compounds persists at higher-pressure level when the (S) thermal field is accounted for. The dephasing between the two thermal field calculations which was observed at 10$^{-2}$\,mbar is now enhanced here. C$_2$H$_6$ is more abundant at the equator using the (S) field, and has now a steeper abundance gradient toward the South pole. The temperature of the (U) field corresponds to temperatures of the (S) field at 40$^{\circ}$S during the winter solstice.

Note that the resolution of the model in the pressure space varies with thermodynamic conditions as discussed in \ref{subsubsection:Seasonal_T_map}, while the vertical resolution of the model is constant. The altitudinal resolution of the model has been doubled in order to assess if the differences observed between the two thermal fields at high latitudes in the winter hemisphere where not due to numerical artifact. The results obtained were identical.

\begin{figure}[htp]
\begin{center}
\includegraphics[width=0.6\columnwidth]{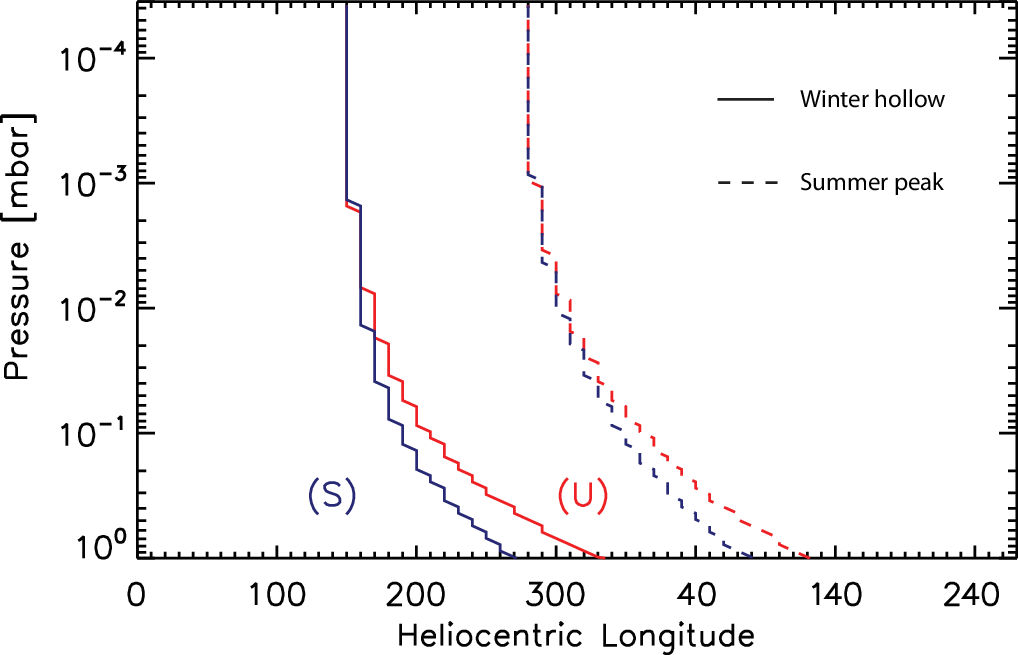}
\caption{Evolution of the seasonal maximum value (\textit{summer peak}, dashed lines) and minimum value (\textit{winter hollow}, solid lines) reached by the C$_2$H$_6$ mole fraction as a function of pressure at 80$^{\circ}$S. Calculations using the (S) and the (U) thermal fields are denoted by blue and red colors, respectively.}
\label{fig:TimeShift}
\end{center}
\end{figure}

It is clear from Figs \ref{fig:Seasonnal_Comparaison_2} and \ref{fig:Seasonnal_Comparaison_3} that accounting for the (S) thermal field leads to a decrease in the seasonal lag of C$_2$H$_6$ and C$_2$H$_2$ at pressure levels higher than 10$^{-2}$\,mbar. We also noted that these two compounds still show seasonal variability at 1\,mbar while this variability has already vanished for C$_2$H$_6$ with the (U) thermal field. The temporal positions of the  maximum and minimum abundance values as a function of pressure, corresponding to summer and winter conditions (hereafter called summer peak and winter hollow, respectively), are displayed in Fig. \ref{fig:TimeShift}. We note an increase in the phase lag between the (U) and the (S) thermal field calculations with increasing pressure from the 10$ ^{-2}$\,mbar to the 1\,mbar pressure level. At 1\,mbar, the depletion in the C$_2$H$_6$ abundance due to low-insolation winter conditions (winter hollow) is occurring 60$^{\circ}$ in heliocentric longitude earlier with the (S) thermal field than with the (U) thermal field. Similarly, the increase in C$_2$H$_6$ abundance due to summer conditions (summer peak) is occurring 40$^{\circ}$ in heliocentric longitude earlier.

We can note here some discrepancies between both our study cases and the following statement previously made by \citet{Moses2005b}:
\textit{"Our assumption of a constant thermal structure with time and latitude introduce mole-fractions errors of a few percent but will not affect our overall conclusions."}
This statement seems to be clearly in disagreement with the results presented in Figs. \ref{fig:Seasonnal_Comparaison_1}, \ref{fig:Seasonnal_Comparaison_2} and \ref{fig:Seasonnal_Comparaison_3} where the differences between the (U) and (S) cases are well beyond a few percent. The fact that we did not reach the same conclusions lies in the different approach we had.
In the present model, we assumed that the compounds mole fractions followed the atmospheric contraction/dilatation in the pressure space with changing thermodynamic conditions (see \ref{subsubsection:Seasonal_Temp_Use}). Therefore, when the atmosphere column contracts, the mole fraction vertical gradients in altitude are increased and the diffusion to higher pressure levels becomes faster. On the other hand, when the thermal structure does not evolve with time (i.e. the (U) study case), our approach is identical to the work of \citet{Moses2005b}.

The diffusion timescale is therefore decreased at all times in the (S) model with respect to the (U) model, due to the thermal evolution of the (S) model. Consequently, this decrease in the diffusion timescale shifts to higher pressures the level where the seasonal variations vanish. This effect is maximized at the poles where the seasonal variations in temperature are important.

We remind the reader that we assumed a seasonally and latitudinally constant eddy diffusion profile, due to the lack of constraint on this free-parameter.

\section{Comparison with Cassini/CIRS data}
\label{section:Comparision_CIRS}

After Cassini's arrival in the Saturn system, observations of hydrocarbons have been performed with an unprecedented spatial and temporal coverage, either with nadir \citep{Howett2007, Hesman2009, Sinclair2013} or limb \citep{Guerlet2009, Guerlet2010} observing geometries. \citet{Howett2007} reported the meridional variability of C$_2$H$_6$ and C$_2$H$_2$ from 15$^{\circ}$S to almost 70$^{\circ}$S at a pressure level around 2\,mbar between June 2004 ($L_s$ $\approx$ 292$^{\circ}$) and November 2004 ($L_s$ $\approx$ 298$^{\circ}$) with Cassini/CIRS. The C$_2$H$_2$ distribution was found to peak around 30$^{\circ}$S and decreases towards both the equator and the South Pole. C$_2$H$_6$ showed a rather different and puzzling behavior, with an increasing abundance southward from the equator, confirming the earlier findings of \citet{Greathouse2005} and \citet{SimonMiller2005}.

\citet{Sinclair2013} reported observations of C$_2$H$_6$ and C$_2$H$_2$ around the 2.1\,mbar pressure level, acquired in nadir observing mode with a good spatial coverage, from South to North Pole. These observations range in time from March 2005 ($L_s$ $\approx$ 307$^{\circ}$) to September 2012 ($L_s$ $\approx$ 37$^{\circ}$). However, the most recent ones were contaminated with the signature of Saturn's 2011 Great Storm \citep{Fletcher2011,Fischer2011,SanchezLavega2011}, which stratospheric aftermath was studied extensively by \citet{Fletcher2012}. Their retrieval suggests that C$_2$H$_2$ is abundant at the equator and decreases toward the poles. Similarly to previous findings, C$_2$H$_6$ showed a behavior different from C$_2$H$_2$. The observed trend suggests an enrichment in C$_2$H$_6$ at high southern latitudes.

\begin{figure}[htp]
\begin{center}
\includegraphics[width=1.0\columnwidth]{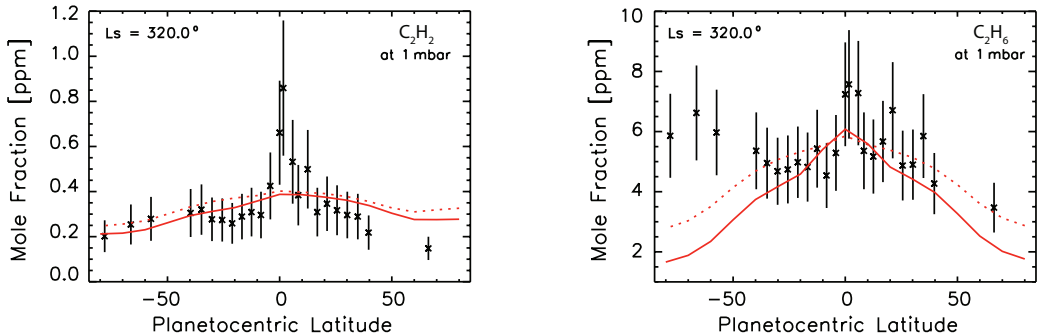}
\caption{Comparison between the Cassini/CIRS limb observations \citep{Guerlet2009} for heliocentric longitudes ranging from $L_s$ $\approx$ 300$^{\circ}$ to $L_s$ $\approx$ 340 $^{\circ}$ and the photochemical model predictions. Photochemical predictions are presented for a heliocentric longitude of 320$^{\circ}$. C$_2$H$_2$ and C$_2$H$_6$ are respectively shown in the left and right panels. Model outputs using the (S) and (U) thermal fields are denoted by solid and dotted lines, respectively. No rescaling factors have been applied, see text for details.}
\label{fig:Guerlet_CIRS1}
\end{center}
\end{figure}

The observations retrieved by \citet{Guerlet2009,Guerlet2010} from Cassini/CIRS limb-scans, enabled constraining the vertical distributions of C$_2$H$_2$ and C$_2$H$_6$ from 5\,mbar to 5 $\mu$bar. The retrieved mole fraction meridional profiles of C$_2$H$_2$ and C$_2$H$_6$ at 1\,mbar are presented in Fig. \ref{fig:Guerlet_CIRS1} for a better comparison with already published nadir observations \citep{Howett2007,Sinclair2013}. When considering the meridional profiles only, the systematic errors are not considered and therefore the observational errors have been reduced of 20\,\% in Fig. \ref{fig:Guerlet_CIRS1}. We chose not to rescale our predictions to superimpose them to the observations, although it is occasionally observed in the literature. \citet{Dobrijevic2003} and \citet{Dobrijevic2010a} showed that uncertainty propagations in giant planet photochemistry lead to uncertainties of about an order of magnitude in the C2 species' abundances at arround the 1\,mbar pressure level. Recent improvement on the chemical scheme greatly reduced these uncertainties \citep{Hebrard2013,Dobrijevic2014,Loison2014} (see Fig. \ref{fig:Reduction}) by a factor of 1.8 for C$_2$H$_6$ and a factor of 4.2 for C$_2$H$_2$. The absolute differences between the photochemical predictions and the observations shall not be seen as a concern if it remains within the photochemical uncertainties. What can (and have to) be compared between observations and models are the general trends seen in the meridional distributions at the relevant pressure levels.

\begin{figure}[htp]
\begin{center}
\includegraphics[width=1.0\columnwidth]{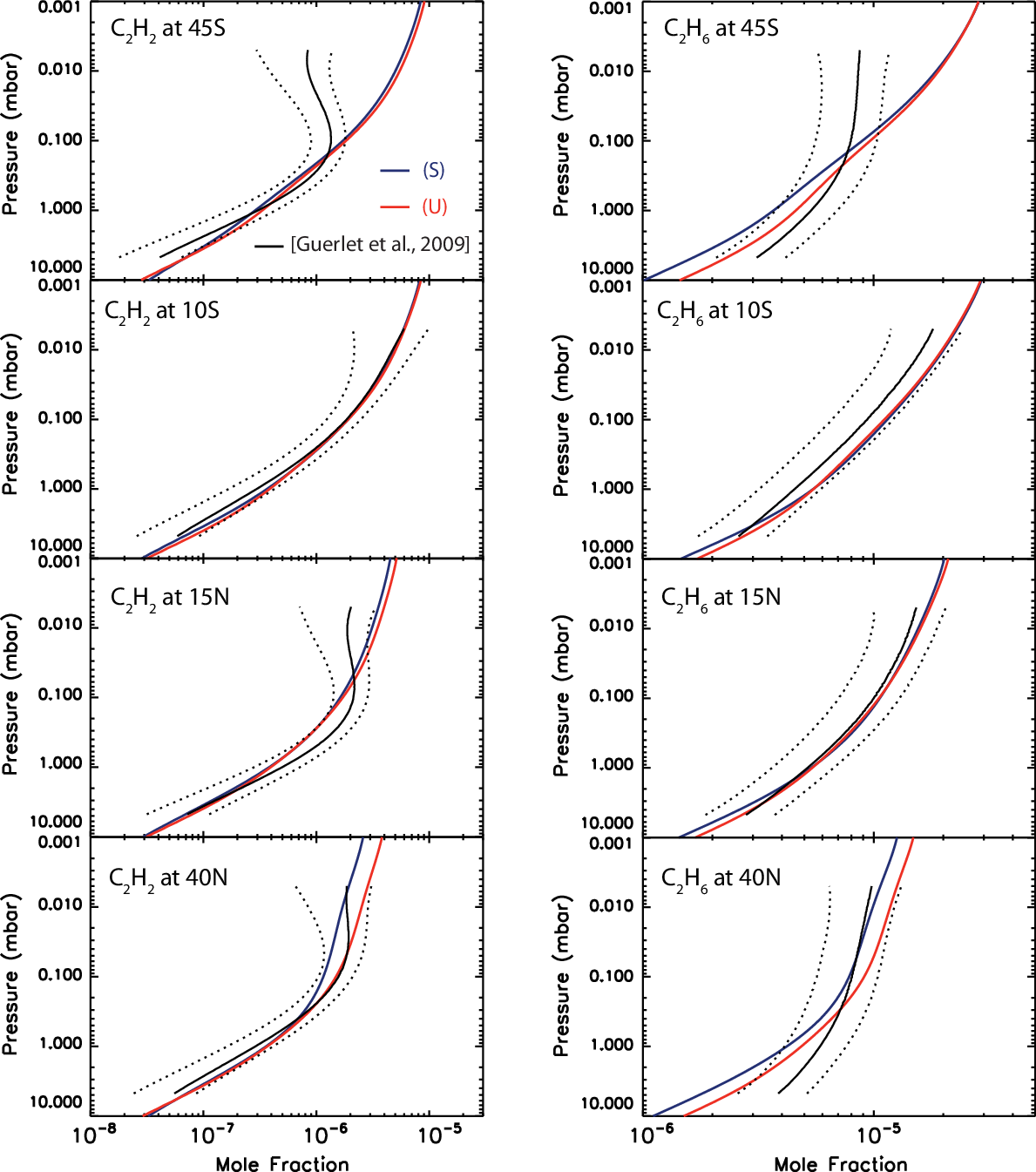}
\caption{Comparison between the C$_2$H$_2$ (left panels) and C$_2$H$_6$ (right panel) retrieved abundances with the photochemical predictions over the pressure sensitivity range at few observed planetographic latitudes. Photochemical predictions that uses the (S) and (U) thermal field are displayed by solid blue and red colors, respectively. Solid and dotted black lines represent the observed abundances of \citet{Guerlet2009} with the 1-$\sigma$ uncertainties, respectively. Photochemical predictions are presented for a heliocentric longitude of 320$^{\circ}$.}
\label{fig:Guerlet_CIRS2}
\end{center}
\end{figure}

\begin{figure}[htp]
\begin{center}
\includegraphics[width=0.8\columnwidth]{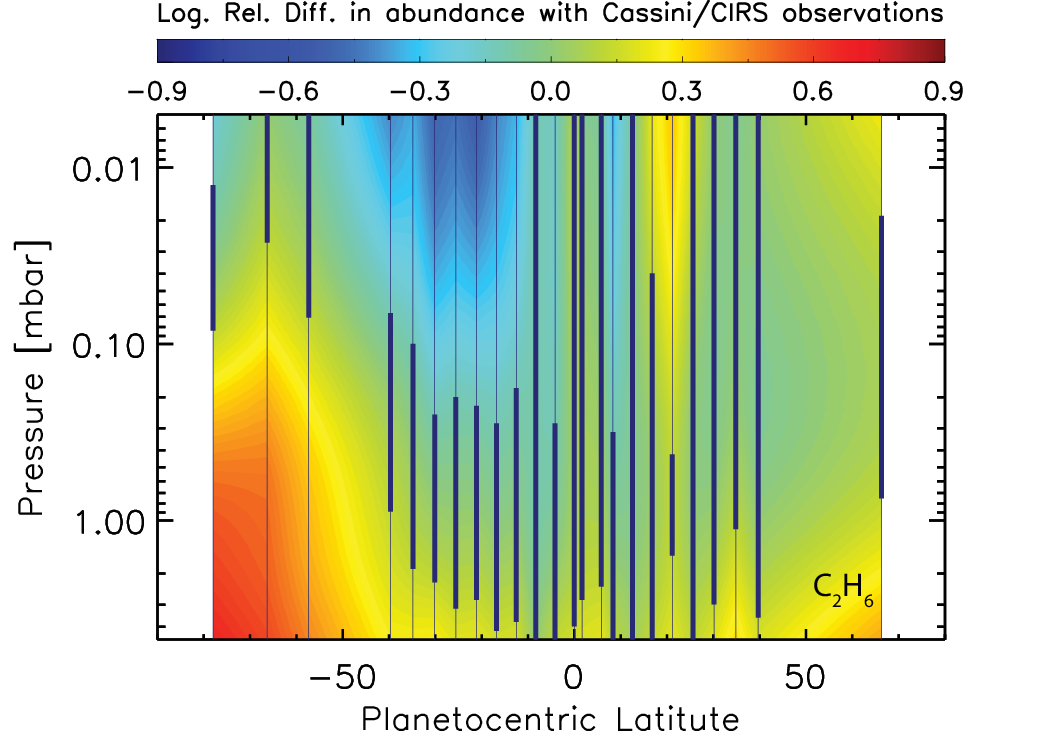}\\
\includegraphics[width=0.8\columnwidth]{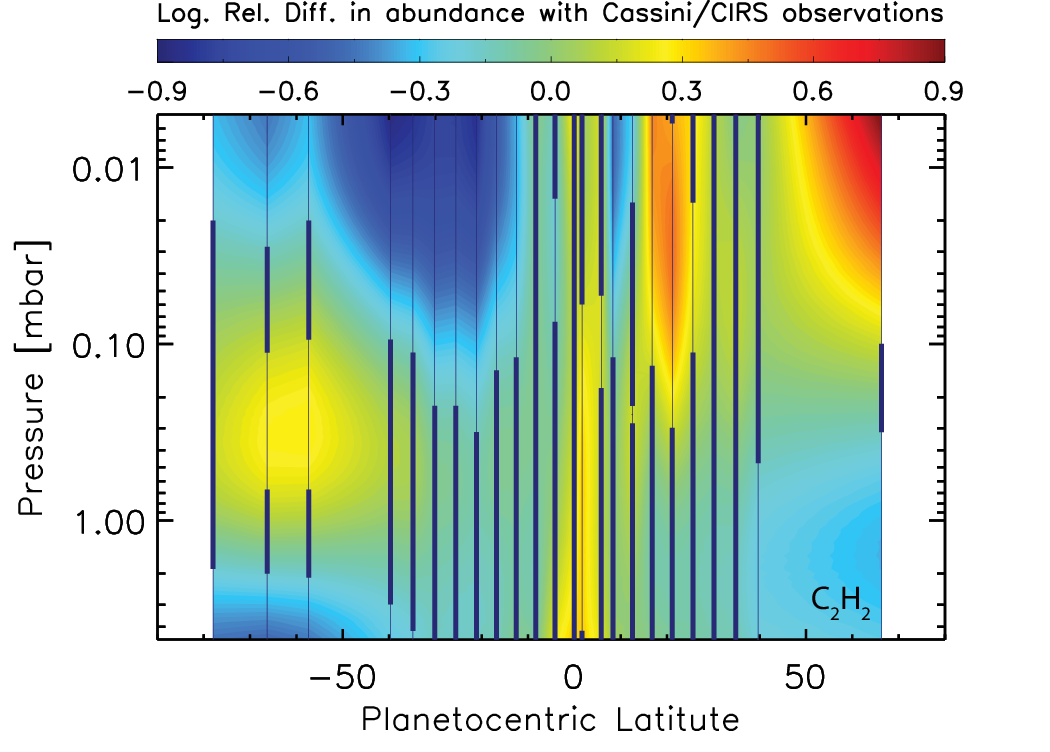}
\caption{Comparison between observations \citep{Guerlet2009} and photochemical predictions as a function of the pressure sensitivity range of limb observations and planetocentric latitudes. C$_2$H$_6$ and C$_2$H$_2$ are presented in the upper and lower panel, respectively. The observing period ranges from L$_S$ = 300$^{\circ}$ and L$_S$ = 340$^{\circ}$. Observations are compared to photochemical predictions at L$_S$ = 320$^{\circ}$. The logarithm of the difference between Cassini observations of these compounds and the photochemical predictions that use (S) thermal field is plotted here. Positive/negative values denote an under/over-prediction of the photochemical models. The vertical lines denote the latitudes for which observations have been made. The thick portions of those lines show the region where the photochemical predictions are within the observation uncertainties. Uncertainties on the photochemical predictions are not taken into account. No scaling factor have been applied.}
\label{fig:2D_contour}
\end{center}
\end{figure}

The Cassini limb-observations offer latitude and altitude information. The retrieved abundances of these two molecules over the pressure sensitivity range are presented in Fig. \ref{fig:Guerlet_CIRS2} at a few observed planetographic latitudes. We have selected these latitudes in order to display the different features noted when confronting these observations with the model, \textit{e.g.} a large over-prediction of their abundance at high southern latitudes and low pressure level, a good agreement at mid-to-low latitudes, and an under-prediction at mid-to-high northern latitudes and high pressure levels, especially noted for C$_2$H$_6$. It is worth presenting a comparison with these data in two-dimensional plots for the sake of clarity. The relative differences between the Cassini observations of C$_2$H$_6$ and C$_2$H$_2$ from the photochemical predictions using (S) thermal field are displayed in Fig. \ref{fig:2D_contour} using a logarithmic scale. The positive and negative values therefore denote the regions where the photochemical model under and over-predicts the abundance of these compounds, respectively. The quantity $log(y_i^{CIRS}/y_i^{PM})$ is plotted as a function of the pressure range and the latitude, where $y_i$ denotes respectively the mole fraction of species $i$, while $PM$ stands for photochemical model. Because the latitudinal coverage of Cassini limb-observations is limited, we have indicated the latitude of each observation by black vertical lines.

\subsection{C$_2$H$_2$}

At the 1\,mbar pressure level (Fig. \ref{fig:Guerlet_CIRS1}), our photochemical model simulations agree reasonably well with the meridional trend seen in the C$_2$H$_2$ meridional profiles, namely the poleward decrease of its abundance, as reported by \citet{Guerlet2009,Guerlet2010} and \citet{Sinclair2013}. Differences are observed in the equatorial regions, at latitudes lower than 15$^{\circ}$ and Northward of 35$^{\circ}$N.

Below the 0.1\,mbar pressure level (see Fig. \ref{fig:2D_contour}), the agreement with the C$_2$H$_2$ distribution reported by \citet{Guerlet2009} is within the uncertainty range of the observations from the equator to $\pm$ 40$^{\circ}$. The agreement is however poor at lower-pressure levels and high-southern latitudes, where C$_2$H$_2$ abundance tends to be over-predicted.

In the equatorial zone, the differences between our photochemical predictions and the observations change sharply over a short latitudinal range. Fig. \ref{fig:Guerlet_CIRS1} shows that our model does not reproduce the equatorial peak of C$_2$H$_2$ abundance (between 5$^{\circ}$S and 5$^{\circ}$N roughly). This peak is thought to be caused by Saturn's thermal Semi-Annual Oscillation (SSAO) \citep{Orton2008,Fouchet2008,Guerlet2009,Guerlet2011}.

In the southern hemisphere, from 10$^{\circ}$S to 40$^{\circ}$S, we over-predict the C$_2$H$_2$ mole fraction at pressures lower than 0.1\,mbar. This feature is also observed when comparing C$_2$H$_6$ predictions to observations and is discussed below.

\subsection{C$_2$H$_6$}

\citet{Moses2005b} have shown that the seasonal variability of C$_2$H$_6$ at pressures higher than 0.8\,mbar was negligible, because the timescale driving this compound's abundance becomes longer than the Saturn year below that pressure level. 

Due to the larger chemical evolution timescale of C$_2$H$_6$ with respect to C$_2$H$_2$, this compound is expected to be more sensitive to transport processes than the latter one. In addition to that, the uncertainties on the reactions involved in the production or destruction of C$_2$H$_2$ lead to an important error bar on its vertical profile (recall Fig. \ref{fig:Reduction}), whereas the predicted shape of C$_2$H$_6$ has smaller error bars. Therefore, C$_2$H$_6$ is a compound that should be first used to trace dynamics rather than C$_2$H$_2$. C$_2$H$_6$ is reasonably well reproduced from the equator to 40$^{\circ}$ in both hemispheres below the 0.1\,mbar pressure level. An equatorial peak similar to the one observed in the C$_2$H$_2$ meridional profile (see Fig. \ref{fig:Guerlet_CIRS1}), though with a smaller relative amplitude, is present in the C$_2$H$_6$ meridional profile and could also be caused by the SSAO. Similarly to C$_2$H$_2$, we underpredict C$_2$H$_6$ abundance from 40$^{\circ}$S to south pole below the 0.1\,mbar pressure level, and we overpredict its abundance at latitudes ranging from 10$^{\circ}$S to 40$^{\circ}$S above the 0.1\,mbar pressure level. These similar over/underprediction seen in C$_2$H$_2$ and C$_2$H$_6$ could be caused by large scale dynamical cell redistributing species meridionally in Saturn's stratosphere, as suggested by \citet{Guerlet2009, Guerlet2010}, \citet{Sinclair2013}.

At the 1\,mbar pressure level (Fig. \ref{fig:Guerlet_CIRS1}), the photochemical predictions that use the (S) thermal field predict a steeper equator-to-pole gradient, which arise from the faster diffusion to higher-pressure levels when considering the (S) thermal field. Fitting these meridional profiles with dynamical processes could help to constrain meridional mixing processes and will be the object of a forthcoming study.


\subsection{Does accounting for the seasonal evolution of the thermal field better fit Cassini data?}

\begin{figure}[htp]
\begin{center}
\includegraphics[width=0.8\columnwidth]{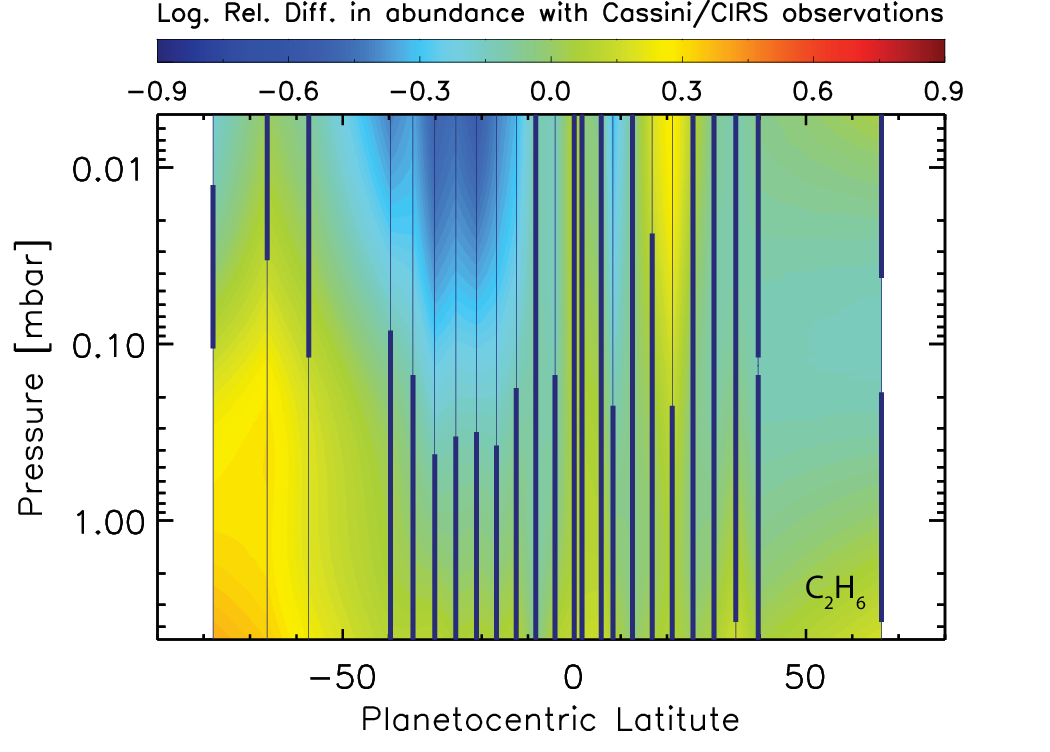}\\
\includegraphics[width=0.8\columnwidth]{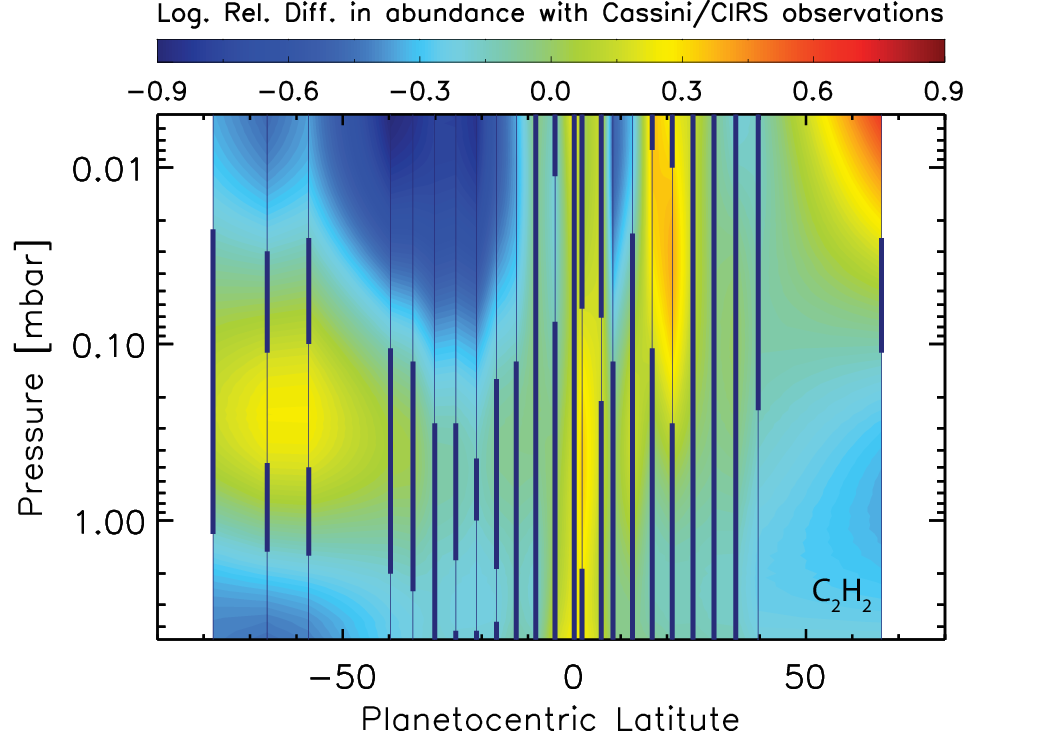}
\caption{Same as Fig. (\ref{fig:2D_contour}) using the photochemical predictions that use (U) thermal field.}
\label{fig:2D_contour_2}
\end{center}
\end{figure}

Fig. \ref{fig:2D_contour_2} presents the comparison between Cassini-limb observations with photochemical predictions using the (U) thermal field in a similar way as Fig. \ref{fig:2D_contour}. C$_2$H$_6$ is better predicted using (U) thermal field below 1\,mbar while C$_2$H$_2$ is better predicted at all pressure levels using (S) thermal field. The region where C$_2$H$_6$ was widely under-predicted with the (S) thermal field from mid-southern latitude to South pole (recall Fig. \ref{fig:2D_contour}), is now sligthly reduced, but it remains significant.

\begin{figure}[htp]
\begin{center}
\includegraphics[width=1.0\columnwidth]{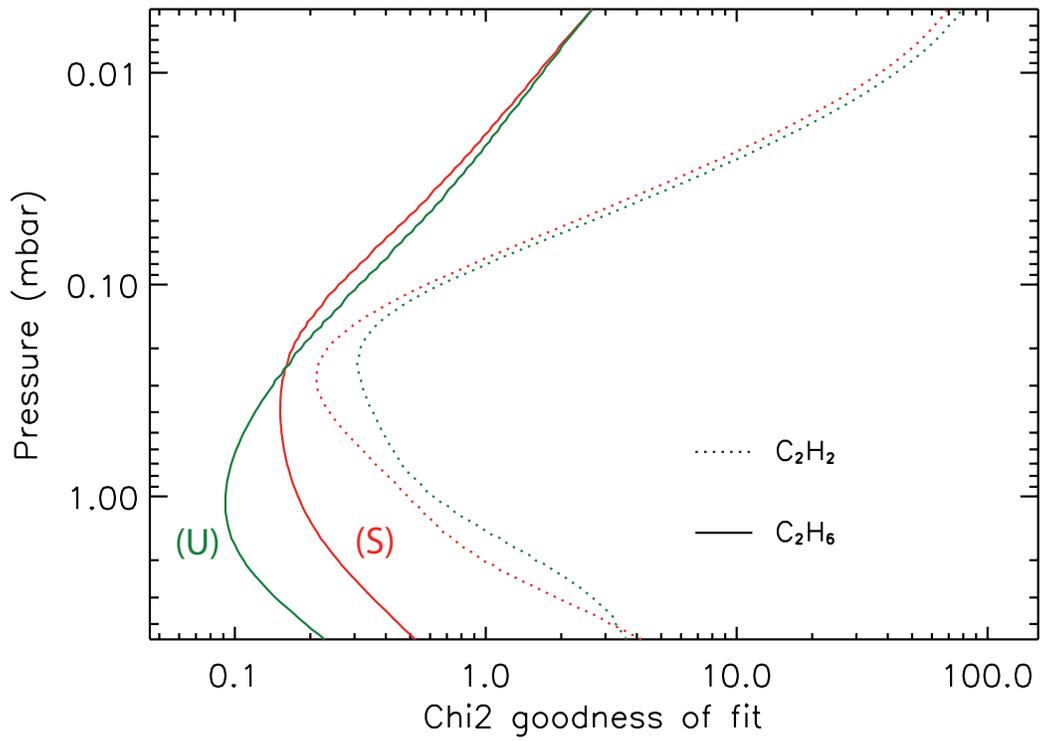}
\caption{Evolution of the $\chi$ square goodness-of-fit factor for C$_2$H$_2$ (dotted lines) and C$_2$H$_6$ (dashed lines) over the sensitivity pressure ranges of the Cassini/CIRS limb observation mode. This factor is presented when using (S) and (U) thermal fields, denoted by the red and green lines respectively. This factor is summed for all observed latitudes.}
\label{fig:Chi_2_Guerlet}
\end{center}
\end{figure}

The evolution of the Chi-Square goodness-of-fit between the predicted C$_2$H$_2$ and C$_2$H$_6$ abundance (using both thermal fields) and Cassini observations for every observed latitudes is shown in Fig. \ref{fig:Chi_2_Guerlet}. These values are computed first by interpolation of the photochemical prediction on the observed latitudes and pressure levels, then the Chi-Square coefficient presented in Fig. \ref{fig:Chi_2_Guerlet} is computed and summed over the observed latitudinal range. Using the (S) thermal field in the predicted C$_2$H$_6$ profile represents a slight improvement at pressures lower than 0.2\,mbar. However, in the lower stratosphere, from 0.2\,mbar to 5\,mbar, the predicted C$_2$H$_6$ shape is better reproduced using the (U) thermal field. C$_2$H$_2$ is better reproduced using the (S) thermal field at all pressure levels.

\subsection{Discussion}

As noted above, it has been shown that both C$_2$H$_6$ and C$_2$H$_2$ were overpredicted at mid-southern latitudes and above 0.1\,mbar, while C$_2$H$_6$ was underpredicted at high-southern latitudes and below 0.01\,mbar (recall Fig. \ref{fig:2D_contour}). It can be pointed out that the eddy diffusion coefficient used in this work was possibly not the most optimal one. We remind the reader that the coefficient used here was chosen so that it provides a satisfactory fit of the CH$_4$ vertical profile in comparison to Voyager/UVS and Cassini/CIRS data \citep{Smith1983,Dob2011}.

However, when looking carefully at the southern mid latitudes in Figs. \ref{fig:Guerlet_CIRS2} and \ref{fig:2D_contour}, where we overpredict the C$_2$H$_6$ abundance above 0.1\,mbar and underpredict its abundance below 1\,mbar, a diffusion coefficient greater in the upper stratosphere (above 0.1\,mbar) and smaller below that level might provide a better fit at these latitudes. We have performed a sensitivity study on that parameter by using several dozen different eddy diffusion coefficients. The seasonal model was run from the converged state with the nominal eddy diffusion profile presented in this work. From that point, the eddy diffusion profile was modified and several additional iterations over Saturn orbits were necessary for the system to converge, depending on the new eddy diffusion coefficient. Several different eddy diffusion coefficients provide a better fit of mid-latitudes in the southern hemisphere, but at the same time the good agreement in the northern hemisphere vanishes.

All along this study, we have assumed an eddy coefficient constant with latitudes. However, the sensitivity study on that free parameter has shown us that some latitudes that were not properly reproduced with an eddy diffusion coefficient based on globally averaged observations could be reproduced when adjusting this coefficient. In principle, it could therefore be possible to fit all the observed latitudes by finding the most adapted diffusion coefficient at each latitude. We have not explored this possibility more extensively since the meridional variability of that parameter has not yet been demonstrated with the observation of the latitudinal variability of CH$_4$ homopause.

Another likely possibility would be the existence of large scale meridional transport (diffusive and/or advective), as the associated atmospheric motions might be an important source of departure from the photochemical predictions \citep{Moses2005b,Guerlet2009, Guerlet2010, Friedson2012, Sinclair2013}. However, it is difficult to assess the validity of this hypothesis without further numerical testing. Such work is defered to a later paper. The recommended methodology would be first to retrieve the latitudinal variability of the CH$_4$ homopause in order to better constrain the latitudinal variability of the eddy diffusion coefficient before fitting the Cassini/CIRS observations with meridional transport.


Predictions of the existence of large-scale circulation have already been studied using numerical models. \citet{Conrath1990}, using a 2D zonally averaged model, have predicted a summer-to-winter pole stratospheric circulation cell for Saturn at solstice with upwelling around the summer pole and downwelling around the winter pole. They also found a double circulation cell at equinox, with upwelling at equator and downwelling at both poles. However their calculations did not include aerosols heating which is known to impact the radiative budget in the atmosphere. Afterwards, \citet{West1992} included such heating for a Jovian-like planet and showed that circulation was altered above 100\,mbar, with downwelling at equator and upwelling at the poles. We note that they have not presented the effect of aerosol heating on a Saturn-like planet at solstice. Recently, \citet{Friedson2012} predicted a seasonally reversing circulation cell using a 3D GCM, with upwelling at the equator and downwelling at low latitudes in the winter hemisphere. The next step of this study is to evaluate whether these predicted circulation patterns are sufficient for our full 2D photochemical model to reproduce the Cassini data.

\section{Summary and perspectives}
\label{section:Summary}

We have developed a latitudinally and seasonally variable photochemical model for giant planets and we have adapted it first to Saturn. The model takes into account photochemistry, vertical mixing, Saturn's obliquity, and variation of seasonally dependent orbital parameters such as subsolar latitude, heliocentric distance and ring shadowing. Meridional transport (both advective and diffusive) as well as vertical advective transport have been coded in the model but have been switched off for the current study. The present model is therefore run as a sum of 1D seasonally variable models calculated at different latitudes. This is the first step toward a full-2D model as it is already coupled to a full-3D radiative transfer model. This first paper is dedicated to the study of the photochemical effects involved by a seasonally variable thermal field, which relies on predictions from a radiative climate model \citep{Greathouse2008}.


\textit{Seasonal variability of C$_2$H$_2$ and C$_2$H$_6$.} - The seasonal variations of the C2-hydrocarbon mole fractions such as C$_2$H$_2$ (acetylene) and C$_2$H$_6$ (ethane) are important at pressure levels lower than 0.1\,mbar and at high latitudes. These compounds are known to act as the main stratospheric coolants \citep{Yelle2001} and we therefore have emphasized them in this work on their seasonal variabilities. These hydrocarbons are produced by chemical reactions involving radicals which strongly depend on insolation. Including the ring shadowing effect generally results in lowering the mole fraction of hydrocarbons at latitudes that are under the shadow of the rings. The decrease in the mole fraction of these compounds caused by the rings is more important for C$_2$H$_2$ than for C$_2$H$_6$. This decrease tends to vanish at higher-pressure levels.

\textit{Accounting for a seasonally variable thermal field.} - We found that including a seasonally variable thermal field mainly impacts the seasonal evolution of the hydrocarbon mole fractions in two ways. First, the modification of the thermal field amplitude will affect the position of the methane homopause and will impact consequently the integrated production rates of radicals above the pressure level of 10$^{-4}$\,mbar. Then, accounting for a seasonally variable thermal field will affect the diffusion of these produced hydrocarbons through the contraction and dilatation of the atmospheric columns. This will compress the vertical distribution of the atmospheric compounds and will therefore increase the diffusion of these compounds to higher pressure levels. Because the former effect is correlated with the seasonal variation of temperature of the atmospheric column, it is therefore increased with increasing latitude, i.e. where the seasonal thermal gradients are strong. At 10$^{-4}$\,mbar, C$_2$H$_2$ and C$_2$H$_6$ seasonal abundance gradients are generally enhanced. At this pressure level and during the summer season, C$_2$H$_6$ shows a positive equator-to-summer pole (North or South) gradient. At 10$^{-2}$\,mbar, the seasonal abundance gradients are also increased: the abundance of C$_2$H$_6$ and C$_2$H$_2$ are respectively decreased by a factor of 2.2 and 1.7 at a latitude of 80$^{\circ}$ during winter, with respect to the (U) study case.
We do not reach the same conclusions than \citet{Moses2005b} about the thermal sensitivity of the photochemical model. While our (U) study case is similar to their approach, our (S) study case is, on the other hand, different in a way that the compounds mole fractions are assumed to follow the atmospheric contraction/dilatation in the pressure space with changing thermodynamic conditions. This consequently affects the downward diffusion of the seasonally produced photochemical by-products.


\textit{Comparison with Cassini/CIRS data.} - The Cassini spacecraft has now provided an important amount of data that includes good spatial and temporal coverage \citep{Sinclair2013} as well as good vertical sensitivity \citep{Guerlet2009,Guerlet2010}. Our model reproduces reasonnably well the meridional distributions of C$_2$H$_2$ and C$_2$H$_6$ up to mid-latitudes, even without meridional circulation. However, the overall increase of C$_2$H$_6$ from the south pole towards the north pole at L$_s$=320$^{\circ}$ is not reproduced. An interesting feature has been noted: our model tends to underpredict C$_2$H$_6$ abundance from 40$^{\circ}$S to South pole at pressure levels ranging from 5\,mbar to 0.1\,mbar and overpredict its abundance at latitude ranging from 10$^{\circ}$S to 40$^{\circ}$S at pressure ranging from 5\,$\times$\,10$^{-3}$ to 0.1\,mbar. These results are consistent with previous findings of \citet{Guerlet2009, Guerlet2010} and \citet{Sinclair2013}. This interesting feature is also observed for C$_2$H$_2$, although in a less pronounced way. The forthcoming step is to turn on the meridional transport to evaluate if it can help better fit to the Cassini data. 


\textit{Coupling radiative climate model and photochemical model} Accounting for these results may have important implications for radiative climate models and GCMs because the predicted temperatures from these models are very sensitive to the amount of these coolants \citep{Greathouse2008}. From 0.1 to 10$^{-4}$\,mbar, where their seasonal variability is important, the increase in the amount of these coolants during the summer season will likely counteract the increase in the atmospheric heating caused by the increase in the solar insolation at high-latitudes. Depending on the relative magnitudes of the photochemical timescale over the thermal inertia timescale, the peak in the predicted temperatures at high-latitudes could happen earlier around the summer solstice than previously predicted using a radiative climate model which uses time-independant abundances of atmospheric coolants. The predicted temperatures are therefore expected to start decreasing earlier after summer solstice than what would be predicted with a model that held the amount of these atmospheric coolants constant over time. Moreover, the effect on the lower-stratosphere could be also interesting, as we showed that accounting for the seasonal evolution of the thermal field impacts the phase-lag and the seasonal variability in this region. 


\section{Acknowledgement}

This work has been supported by the \textit{Investissements d'avenir} program from \textit{Agence Nationale de la Recherche} under the reference ANR-10-IDEX-03-02 (IdEx Bordeaux). Part of this work was done by V.H. at Southwest Research Institute, San Antonio. We thank Julianne Moses and an anonymous reviewer for their constructive comments on the manuscrit. We thank Jean Brillet for providing his line-by-line radiative transfer model as well as for useful discussions. We thank Sandrine Guerlet for providing their published Cassini data and for useful discussions. We acknowledge Eric H\'ebrard for useful discussions on chemistry, Franck Selsis for interesting comments on photochemical models, James Sinclair, Christophe Cossou and Marcelino Agundez for useful comments and discussions. We thank Aymeric Spiga and Melody Sylvestre for useful discussions.

\bibliographystyle{elsarticle-harv}
\bibliography{Draft}

%

\end{document}